\documentclass[sigconf]{acmart}

\usepackage[ruled,vlined]{algorithm2e}
\usepackage{algorithmic}
\usepackage{subfigure}
\usepackage{graphicx}
\usepackage{multirow}
\usepackage{array}
\usepackage{colortbl}
\usepackage{booktabs} \usepackage{enumitem}
\usepackage{xcolor}
\usepackage{amsmath}
\usepackage{bm}
\usepackage{amsthm}
\usepackage{balance}

\definecolor{Ocean}{RGB}{173, 216, 230}

\definecolor{Lace}{RGB}{253 ,245,230	}

\newtheorem{theorem}{Theorem}
\newtheorem{lemma}{Lemma}

\AtBeginDocument{%
  }

\copyrightyear{2024}
\acmYear{2024}
\setcopyright{acmlicensed}
\acmConference[CIKM '24]{Proceedings of the 33rd
ACM International Conference on Information and Knowledge
Management}{October 21--25, 2024}{Boise, ID, USA}
\acmBooktitle{Proceedings of the 33rd ACM International Conference on
Information and Knowledge Management (CIKM '24), October 21--25, 2024,
Boise, ID, USA}
\acmDOI{10.1145/3627673.3679590}
\acmISBN{979-8-4007-0436-9/24/10}


\begin{document}


\title{Guaranteeing Accuracy and Fairness under Fluctuating User Traffic: A Bankruptcy-Inspired Re-ranking Approach}

\author{Xiaopeng Ye}
\affiliation{%
\institution{Gaoling School of Artificial Intelligence\\Renmin University of China}
 \city{Beijing}\country{China}
}
\email{xpye@ruc.edu.cn}

\author{Chen Xu}
\affiliation{%
\institution{Gaoling School of Artificial Intelligence\\Renmin University of China}
 \city{Beijing}\country{China}
}
\email{xc_chen@ruc.edu.cn}

\author{Jun Xu}
\authornote{Corresponding author}
\affiliation{%
\institution{Gaoling School of Artificial Intelligence\\Renmin University of China}
 \city{Beijing}\country{China}
}
\email{junxu@ruc.edu.cn}

\author{Xuyang Xie}
\affiliation{%
 \institution{Huawei Noah's Ark Lab}
  \city{Shenzhen}
  \country{China}
 }
\email{xiexuyang@huawei.com}

\author{Gang Wang}
\affiliation{%
 \institution{Huawei Noah's Ark Lab}
  \city{Shenzhen}
  \country{China}
 }
\email{wanggang110@huawei.com}

\author{Zhenhua Dong}
\affiliation{%
 \institution{Huawei Noah's Ark Lab}
  \city{Shenzhen}
  \country{China}
 }
\email{dongzhenhua@huawei.com}

\renewcommand{\shortauthors}{Xiaopeng Ye, et al.}

\begin{abstract}

Out of sustainable and economical considerations, two-sided recommendation platforms must satisfy the needs of both users and providers. Previous studies often show that the two sides' needs show different urgency: providers need a relatively long-term exposure demand while users want more short-term and accurate service. However, our empirical study reveals that previous methods for trading off fairness-accuracy often fail to guarantee long-term fairness and short-term accuracy simultaneously in real applications of fluctuating user traffic. Especially, when user traffic is low, the user experience often drops a lot. Our theoretical analysis also confirms that user traffic is a key factor in such a trade-off problem. How to guarantee accuracy and fairness under fluctuating user traffic remains a problem.
Inspired by the bankruptcy problem in economics, we propose a novel fairness-aware re-ranking approach named BankFair. 
Intuitively, BankFair employs the Talmud rule to leverage periods of abundant user traffic to offset periods of user traffic scarcity, ensuring consistent user service at every period while upholding long-term fairness. Specifically, BankFair consists of two modules: (1) employing the Talmud rule to determine the required fairness degree under varying periods of user traffic; and (2) conducting an online re-ranking algorithm based on the fairness degree determined by the Talmud rule.
Experiments on two real-world recommendation datasets show that BankFair outperforms all baselines regarding accuracy and provider fairness. 
% Experiments on one publicly available and one real industrial dataset show that BankFair outperforms all baselines regarding accuracy and provider fairness. 

\end{abstract}

\maketitle

\section{Introduction}

\begin{figure}[t]
  \centering

\includegraphics[width=1\linewidth]{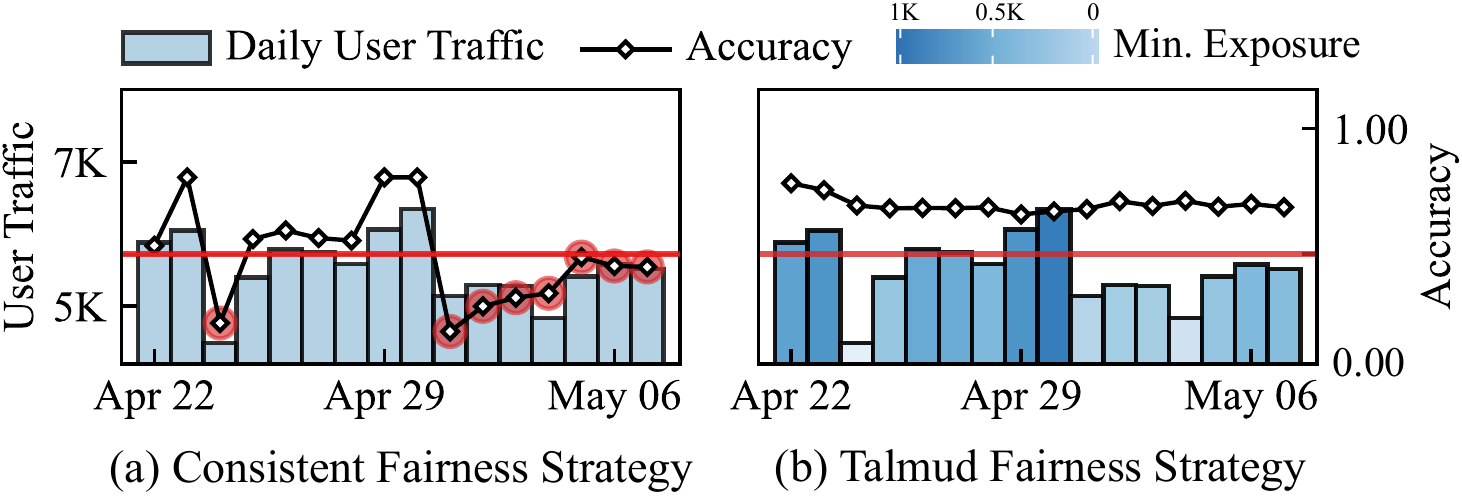}
  \caption{Illustrative experiments based on KuaiRand dataset. (a) Lower user traffic leads to more accuracy loss; (b) Applying distinct fairness strategies based on different traffic levels. }
  \label{fig:intro}

\end{figure}

To create a more equitable and sustainable two-sided platform~\cite{rochet2004two-sided_overview,eisenmann2006strategiesfortwosided,cohen2022competitiontwo-sidedplatforms,wu2021tfrom} (e.g., Amazon and Youtube), recommendation systems (RS) should ensure the needs of both the provider side and the user side~\cite{rochet2004two-sided_overview,wu2021tfrom,patro2020fairrec}. On the provider side, prior studies~\cite{xu2024fairsync,patro2020fairrec,surer2018multistakeholder} have indicated that providers are required to a minimum level of exposure over a period, which aligns with the  ``minimum wage'' policy~\cite{suryahadi2003minimumwage, xu2023p,xu2024fairsync}.
On the user side, established behavior economic theory~\cite{kahneman2013prospecttheory,ge2020userriskpref,paudel2018lossaversion} suggests that users tend to experience increased frustration and dissatisfaction when receiving poor-quality recommendations (i.e., the loss aversion effect~\cite{kahneman2013prospecttheory,kHoszegi2006lossaversionuserexp, paudel2018lossaversion}).
Therefore, platforms should maintain a consistently high standard of recommendation quality to mitigate these negative user experiences~\cite{claridy2009CUSTOMERSERVICE,fleming2006consistency}.

% , since 
% users react more strongly to moments of loss, as described by the loss aversion effect in Prospect Theory~\cite{kahneman2013prospecttheory,ge2020userriskpref,paudel2018lossaversion}.

% established behavior economic theory~\cite{kahneman2013prospecttheory,ge2020userriskpref,paudel2018lossaversion} indicates that 
% users expect to receive recommendations that match their interests and preferences~\cite{pu2011usercentric}, aiming for a positive experience with the recommendation system (RS) during each interaction~\cite{konstan2012user_experience}.
%characterized by amortized fairness

Although the needs of users and providers should be both satisfied, the urgency of the two sides' needs differs.
Providers typically have relatively long-term exposure demands, which do not require immediate fulfillment but rather over an extended period~\cite{wang2023surveyonfairness,xu2024fairsync,yang2023vertical}. 
Conversely, on the user side, the need is more short-term and immediate, because once users receive a poor recommendation, they tend to remember and be impacted by that negative experience for much longer~\cite{kahneman2013prospecttheory,paudel2018lossaversion}.
To balance the needs of both sides, some heuristic methods~\cite{patro2020fairrec,wu2021tfrom,naghiaei2022cpfair} and online gradient descent methods~\cite{xu2023p,xu2024fairsync} have been proposed to trade-off accuracy and fairness.

% existing methods [34 , 51, 52 ] trade-off the accuracy and fairness with different criteria
% enhancing accuracy while ensuring fairness~\cite{xu2023p,surer2018multistakeholder,wu2021tfrom,lopes2024recommendations,patro2020fairrec},optimizing fairness and accuracy together~\cite{xu2023p,zhang2023robust,naghiaei2022cpfair}, amortizing  fairness among users~\cite{biega2018equity,xu2024fairsync,yang2023vertical}.

%that previous fair-aware methods ~\cite{xu2024fairsync,patro2020fairrec,wu2021tfrom} cannot simultaneously satisfy long-term provider fairness and short-term user experience.
%这样写可能会有歧义你的实验到底是在哪个方法上做的

However, the existing trade-off methods could fail in some real applications where the user traffic inevitably fluctuates~\cite {beel2017MrDLib,pires2015youtube}. For example, we conduct a simulation study shown in Figure~\ref{fig:intro}(a) to observe the daily accuracy performances (black line) of a theoretical oracle method under a real industrial dataset KuaiRand~\footnote{\url{https://kuairand.com/}}, which contains 175K user traffic (represented as the blue bars) within half a month.
The red line indicates the minimum daily average accuracy level that the platform can tolerate, below which performance is considered unacceptable.
The oracle method guarantees that each provider receives a predefined minimum exposure each day while maximizing accuracy. 
The result reported in Figure~\ref{fig:intro}(a) shows that the short-term accuracy is more prone to compromise to fulfill long-term fairness requirements, particularly when there are fewer users within a day (red dots).
Similar phenomena have also been observed in most existing fairness-aware models~\cite{xu2024fairsync,wu2021tfrom,patro2020fairrec,wang2023pct,xu2023p}.

To investigate the fundamental reasons behind how user traffic impacts the accuracy-fairness trade-off performances, we provide a theoretical analysis from a constrained optimization problem, as detailed in Section~\ref{sec:harm}. Our analysis reveals that user traffic is a key factor affecting the fairness-accuracy trade-off. Specifically, lower user traffic will result in a more serious accuracy loss when maintaining the same fairness degree (i.e., required minimum exposure) across different periods. Since fluctuations in user traffic are not rare in real-world scenarios~\cite {beel2017MrDLib,pires2015youtube}, a two-sided re-ranking algorithm adaptable to user traffic is needed to balance the needs of both users and providers.

To well incorporate the user traffic into consideration, we utilize the Talmud rule
in bankruptcy problem~\cite{thomson2003axiomatic,curiel1987bankruptcygames} to allocate different fairness requirements across periods of different user traffic. Specifically, the Talmud rule utilizes surplus fairness requirements during periods of high user traffic to compensate for the deficit in fairness requirements during periods of low user traffic, ensuring long-term fairness while enhancing accuracy during periods of low user traffic.
Figure~\ref{fig:intro}(b) shows an illustrative example, where we use the depth of color to represent the fairness degree on that day (i.e., a deeper color of the bar indicates that we will ensure a higher minimum exposure to providers). Across different periods, we utilize the resource-abundant phase (deep color) to compensate for the resource-scarce phase (light color). 
For example, we ensure more minimum exposures to each provider during user-abundant periods (April 26th to April 30th), while reducing the minimum exposure guarantee during user-scarce periods (April 24th, May 1st to May 04th). In such a way, we can amortize the accuracy loss to each day and keep high accuracy levels across all days (i.e., all above the red line), ensuring the needs of both sides.

To implement the above idea, we propose an online re-ranking model called BankFair which can ensure both long-term provider fairness and short-term user accuracy under fluctuating traffic. Specifically, BankFair consists of two modules: (1) Module 1:
we formulate the exposure allocation process as a sequential bankruptcy problem and utilize the Talmud rule to decide the fairness degree of each period; and
(2) Module 2: we utilize the fairness degree obtained from the Talmud rule to guide our online recommendation algorithm, which utilizes the dual method to efficiently solve the fair re-ranking problem in an online style. By combining two modules, we can achieve user-friendly and provider-fair re-ranking effectively and efficiently.

%\xp{To implement the above idea, we proposed our Robust Provider-Fair model, named BankFair, which achieves this robust and fair recommendation through two phases. In the first phase, we assign the provider's requirements to each time step utilizing Talmud rule in bankruptcy problem, which effectively prevents causing significant damage to the ranking accuracy when ensuring a larger minimum exposure for providers during periods of low traffic. In the second phase, we consider the provider-fair re-ranking at each time step. 
%To prevent the constraints from causing too much harm to the objective function, we transform the hard constraint into the adaptive mirror hinge loss penalty and solve it efficiently in the dual space.
%For the online learning process, a stochastic gradient descent is applied to make an effective and efficient online recommendation. Experimental results show that our method can simultaneously ensure long-term fairness for suppliers while maintaining efficient and robust recommendations.

Our main contributions can be summarized as follows:

$\bullet$ We emphasize the significance of guaranteeing both short-term user accuracy and long-term provider fairness under fluctuating user traffic in two-sided platforms.

$\bullet$ We propose a two-sided re-ranking model named BankFair, which formulates the exposure allocation process as a sequential bankruptcy problem and utilizes the Talmud rule to solve it.

$\bullet$ The extensive experiments demonstrate that our approach outperforms existing baselines on two datasets in terms of both fairness and accuracy.

\section{Related Work}

\textbf{Fairness-aware re-ranking in two-sided platforms} has drawn much attention in recent years. Depending on the target audience, it can be divided into three types: user fairness~\cite{li2021user,leonhardt2018userfairness}, provider fairness~\cite{xu2023p,qi2022profairrec,patro2020fairrec, xu2024taxation}, and two-sided fairness~\cite{patro2020fairrec,naghiaei2022cpfair}. For the provider fairness studied in this paper, there are different forms: (1) max-min fairness~\cite{xu2023p}, which aims to ensure the interests of the worst-off providers;
(2) equity of attention~\cite{biega2018equity,wu2021tfrom,naghiaei2022cpfair,morik2020controlling} which lets the exposure received by providers be proportional to their utility; (3) minimum exposure guarantee~\cite{patro2020fairrec,biswas2021fairrecplus,ben2023learning,lopes2024recommendations,xu2024fairsync,surer2018multistakeholder, yang2023vertical}, which tends to ensure that the exposure for providers over a period exceeds a minimum threshold. 
Within the research line, 
some heuristic methods~ \cite{patro2020fairrec,biswas2021fairrecplus,yang2023vertical} were proposed to
ensure the minimum exposure in several recommendation lists and amortize the accuracy loss among each user (e.g., greedy round robin~\cite{patro2020fairrec,biswas2021fairrecplus}).
Also, some other works~\cite{lopes2024recommendations,surer2018multistakeholder,xu2024fairsync} formulated the re-ranking problem with exposure constraint as Integer-Programming (IP) and adopted online optimization methods (e.g., sub-gradient descent~\cite{duchi2011subgradient}) to solve it.
In most existing works~\cite{patro2020fairrec,biswas2021fairrecplus,yang2023vertical,surer2018multistakeholder}, the minimum exposure value cannot be directly specified but is controlled by a hyperparameter, which is hard to adjust and control. 
Thus, FairSync~\cite{xu2024fairsync} considered a more industrially practical setting, which guarantees an arbitrarily specified minimum exposure, and we also consider this scenario in our paper. Nonetheless, all these works overlooked the accuracy loss caused by the minimum exposure fairness under fluctuating user traffic, making it impractical in real-world industrial scenarios.

\textbf{Bankruptcy problem} were first introduced by \citet{o1982problem} and have been widely used in scarce resource allocation scenarios, such as network traffic allocation~\cite{antonopoulos2020bankruptcynetwork}, water resource allocation~\cite{zheng2022water,degefu2018bankruptcywater}, vaccines resource allocation~\cite{hong2021coronavirus}. 
\citet{curiel1987bankruptcygames}, \citet{dagan1993bankruptcycoop} and \citet{thomson2013game} modeled bankruptcy problem as a coalition game, and \citet{curiel1987bankruptcygames} further proved that Talmud solution corresponds to the core of coalition game, which is stable and impregnable. Traditional bankruptcy problem is a one-time allocation process. To adapt to the time-varying nature of RS, we transformed the bankruptcy problem into a sequential version that involves multiple time allocations.

\section{Formulation}
\label{sec:formulation}

We first define some notations for the problem. For vector $\bm{x}\in\mathbb{R}^n$, $\bm{x}_i$ denote the $i$-th element of the vector. For matrix $\bm{X}\in\mathbb{R}^{n\times m}$, let $\bm{X}_{i,j}$ denote the element of $i$-th row and $j$-th column.
$\bm{A}_i$ denote the $i$-th column vector of $\bm{A}$. $\bm{x}\ge \bm{y}$ denotes element $\bm{x}_i$ should be greater or equal to $\bm{y}_i, \forall i$. $k\bm{x}$ denotes that every element $\bm{x}_i$ in $\bm{x}$ will become $k\bm{x}_i$. Next, we give a formal formulation of two-sided re-ranking in RS.

\subsection{Two-sided Re-ranking in Recommender System under Fluctuating User Traffic}

In RS, let $\mathcal{U}, \mathcal{I}, \mathcal{P}$ denote the set of users, items, and providers, respectively. For a provider $p$, there are multiple items in the set $\mathcal{I}_p$ belonging to $p$. When a user $u_t \in \mathcal{U}$ arrives at time $t$, RS will generate a ranking list $L_K^{\text{ori}}(u_t)\in \mathcal{I}^K$ of size $K$. 
Typically, the ranking list can be generated using a user-item relevance vector $\bm{s}_{u_t}\in\mathbb{R}^{|\mathcal{I}|}$, where the relevance score $\mathbf{s}_{u_t, i}\in [0,1]$ is estimated based on the historical information, user profile, etc. 
The goal of two-sided re-ranking is to compute a new list $L_K(u_t)$ which well balances the short-term user accuracy and long-term provider fairness.
When an item $i$ is recommended in $L_K(u_t)$, then the corresponding provider $p$ with such an item (i.e., $i\in\mathcal{I}_p$) can have one exposure.
\begin{figure}[t]
    \centering
\includegraphics[width=1\linewidth]{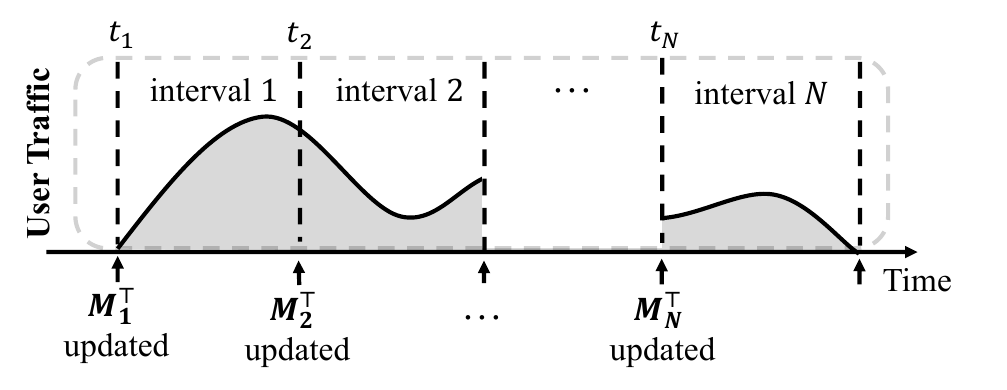}
    \caption{Illustration of algorithm updates over time. The shadow area represents the user traffic of each interval. The minimum exposure needs to be guaranteed at the end of the $N$-th interval.}
\label{fig:update1}
\end{figure}

\subsubsection{Two-sided Re-ranking across Multiple Time Intervals.}
In real scenarios, different users  $\mathcal{U}_n=\{u_t| t_n\leq t <t_{n+1}, n\in [1,2,\ldots,N]\}$ will arrive in RS within time interval $n$, where $t_n$ is the start time of interval $n$. We use $\bm{r}_n=|\mathcal{U}_n|$ to denote the user traffic (i.e., user number) arriving within time interval $n$ (e.g. $n$ means $n$-th hour or $n$-th day during the recommendation process). 

Let $a(n)$ be the average re-ranking accuracy within time interval $n$. Following literature convention~\cite{wu2021tfrom,biega2018equity,xu2023p}, we define the re-ranking accuracy as the ratio between the sum of position-based relevance scores in the re-ranked list $L_K(u_t)$ and that in the original ranking list $L_K^{\text{ori}}(u_t)$, which is detailed in Equation~\eqref{eq:ndcg}. Let $\bm{E}\in \mathbb{N}^{|\mathcal{P}|\times N}$ be the provider earned exposure matrix, where the element $\bm{E}_{p,n}$ denotes the earned exposure for provider $p$ at interval $n$. 

Formally, the two-sided re-ranking task can be written as a linear programming:
\begin{align}
\label{eq:unfied_opt}
f(\phi,\bm{m})=\max_{\bm{E}} &\sum_{n=1}^N a(n;\bm{E}) 
\\ \label{eq:unfied_opt_con1}
        \textrm{s.t.} \quad 
        &  a(n;\bm{E})\ge \phi,\quad \forall n\in [1,2,\ldots,N]\\
% \label{eq:unfied_opt_con2}   
%         &  \bm{E}_{p,n} \ge \bm{M}_{p,n}, \quad \forall p\in\mathcal{P}\\
\label{eq:unfied_opt_con3}
        &  \sum_{n=1}^N \bm{E}_{p,n} \ge  \bm{m}_p, \quad \forall p\in\mathcal{P},
    \end{align}
where $\bm{m}\in \mathbb{R}^{|\mathcal{P}|}$  denote the required minimum exposure and  $\phi \in [0,1]$ is the required minimum accuracy.
The model objective ~\eqref{eq:unfied_opt} aims to maximize the overall re-ranking accuracy among all intervals. The two constraints represent the needs of users and providers respectively:
\begin{itemize}[leftmargin=*]
    \item \textbf{Short-term Accuracy Need}: Constraint~\eqref{eq:unfied_opt_con1} requires the accuracy $a(n)$ at each interval $n$ should be no less than the required minimum accuracy $\phi$.
    \item \textbf{Long-term Exposure Need}: Constraint~\eqref{eq:unfied_opt_con3} demands the cumulative exposure $\sum_{n=1}^N \bm{E}_{p,n}$ of provider $p$ within $N$
intervals should be no less than the required minimum exposure $\bm{m}_p$.

\end{itemize}

\subsubsection{Re-ranking with Algorithm Updates.}
Under real industrial settings as shown in Figure~\ref{fig:update1}, re-ranking model $f(\phi,\bm{m})$ updates its parameters regularly~\cite{zanardi2011dynamicupdating,wei2011differentupdateinterval}.  Under this condition, the ideal re-ranking problem $f(\phi,\bm{m})$ cannot be globally optimized for all intervals. Hence, decomposed sub-problems within each interval will be solved. 

Let $\bm{M}_{p,n}$ be the required minimum exposure for provider $p$ at interval $n$ ($\bm{M}\in\mathbb{R}^{|\mathcal{P}|\times N}$), which will be updated at the beginning of interval $n$, $\forall n=1,2,\cdots, N$.
In the sub-problem, Constraint~\eqref{eq:unfied_opt_con3} is decomposed into two constraint ($\bm{E}_{p,n}\ge \bm{M}_{p,n}, \sum_{n=1}^N\bm{M}_{p,n}\ge\bm{m}_p$).
This requires the summation of the required exposure $\bm{M}_{p,n}$ at each interval should exceed the final required minimum exposure $\bm{m}_p$.
Hence, an optimal $\bm{M}_{p,n}$ should be determined to make $f(\phi,\bm{m})$ is maximized and the accuracy constraint~\eqref{eq:unfied_opt_con1} is satisfied.

\subsection{\mbox{Bankruptcy Problem for Exposure Allocation}}
In this section, we will first introduce the definition and elements of bankruptcy problem. Then we will formulate the exposure allocation task as a bankruptcy problem.
\begin{table}[t]
    \caption{Correspondence between elements in the bankruptcy problem and two-sided re-ranking. }
    \label{tab:compare}
    % \small
    \centering
    \scalebox{0.95}{
    \begin{tabular}{c | c}
    \toprule
Bankruptcy problem & Two-sided re-ranking\\
    \hline agent set $\mathcal{A}$  & time intervals $\mathcal{N}$ \\
 estate $E$ & required minimum exposure $\bm{m}_p$ \\
  claim vector $\bm{d}$ & demanded minimum exposure vector $\bm{D}_{p}$\\
 result vector $\bm{x}$ & predicted minimum exposure vector $\bm{M}_{p}$\\
    \bottomrule
    \end{tabular}  
    }
\end{table}
\subsubsection{Bankruptcy Problem}
Bankruptcy problem provides a resource allocation solution that is suitable for scenarios with fluctuating resources and demands~\cite{thomson2003axiomatic,zheng2022water,antonopoulos2020bankruptcynetwork}.
The input of a bankruptcy problem can be defined as a triplet: $(\mathcal{A}, E, \bm{d})$ and the output can be defined as a result vector $\bm{x}$:

% xujun: 建议Bankrupcy问题不要和fair re-ranking一起说，这样讲不明白。首先介绍完bankrupcy问题，以及它的优化目标。
%然后再把它和fair-ranking中的问题进行对应。
%xujun: 最好总结一张两列的表格，把fair re-ranking中的概念和bankruptcy中的概念总结一下，这样比现在这样的写法清楚。

\textbf{Agent Set} $\mathcal{A}$: the resource demanders which resources need to be allocated to.

\textbf{Estate} $E$: the total resource waited to be allocated $E \in \mathbb{R}$.

\textbf{Claim vector $\bm{d}$}: the demanded resource of agents $\bm{d}\in \mathbb{R}^{|\mathcal{A}|}$.

\textbf{Result vector $\bm{x}$}: the allocation result $\bm{x}\in\mathbb{R}^{|\mathcal{A}|}$, where each $\bm{x}_a$ denotes the allocated resource of agent $a$.
The sum of allocated resources should equal the total resource, i.e., $\sum_{a\in\mathcal{A}}\bm{x}_a=E$.

To solve the bankruptcy problem, for each agent $a\in \mathcal{A}$, the objective is to satisfy the fluctuating agent claims $\bm{d}$ as much as possible when the estate $E$ undergoes fluctuations. 
In economics, several methods have been developed to solve the bankruptcy problem, including the proportional rule~\cite{ertemel2018proportional}, Talmud rule~\cite{thomson2003axiomatic}, etc.

%, Constrained Equal Award Rule, and proportional rule

%In fair re-ranking, the element of $\bm{m}_{p,n}$ can be regarded as the minimum exposures for each provider $p$ for each interval $n$.
% Allocation Vector $x$ & Allocated interval exposure $\hat{\bm{m}}_p$ \\

\subsubsection{Minimum Exposure Allocation as Bankruptcy}

%用一个表格说清楚重要概念间的对应关系
% 好的

We correspond the elements of minimum exposure allocation with those in the bankruptcy problem, as shown in Table~\ref{tab:compare}. Specifically, 

$\bullet$ Time interval $n\in\mathcal{N}$ is regarded as the agent $a\in\mathcal{A}$ in bankruptcy problem. 
This is because $\mathcal{N}$ is the set of intervals we want to allocate the required minimum exposures $\bm{m}$ to.

$\bullet$ The minimum exposure for providers $\bm{m}_p\in \mathbb{R}$ can be regarded as the estate $E$ of the bankruptcy problem, which is the total minimum exposure awaited to be allocated to intervals $\mathcal{N}$.

$\bullet$ The demanded minimum exposure $\bm{D}_p\in \mathbb{R}^{|\mathcal{N}|}$ can be regarded as the claim vector $\bm{d}$ of the bankruptcy problem, which 
stands for the demands of exposure in time intervals.

 $\bullet$ The predicted minimum exposure vector $\bm{M}_p\in \mathbb{R}^{|\mathcal{N}|}$ corresponds to the result vector $\bm{x}$ in the bankruptcy problem, where $\bm{M}_{p,n}$ denotes the actual minimum exposure provider $p$ should receive during time interval $n$.

 Based on our analysis (Section~\ref{sec:harm}), to maintain a high level of accuracy at every interval, we need to set different demands  $\bm{D}_p$ (i.e., the claim $\bm{d}$) based on fluctuating user traffic in real applications~\cite{beel2017MrDLib,pires2015youtube}. Meanwhile, the platform can continuously change its provider-fair policy $\bm{m}_p$ (i.e., the estate $E$) due to the development stage or incentive policy~\cite{bardhan2022more}. Therefore, due to the beneficial characteristics of dynamic resource allocation, we can formulate the exposure allocation process as a bankruptcy problem.

\section{Analysis for Accuracy-Fairness Trade-off}
\label{sec:harm}

In this section, we will analyze the accuracy loss to answer two questions: (1) Why does provider fairness hurt accuracy? (2) Why does the lower user traffic result in accuracy loss?

\begin{figure}
    \centering
    \includegraphics[width=0.95\linewidth]{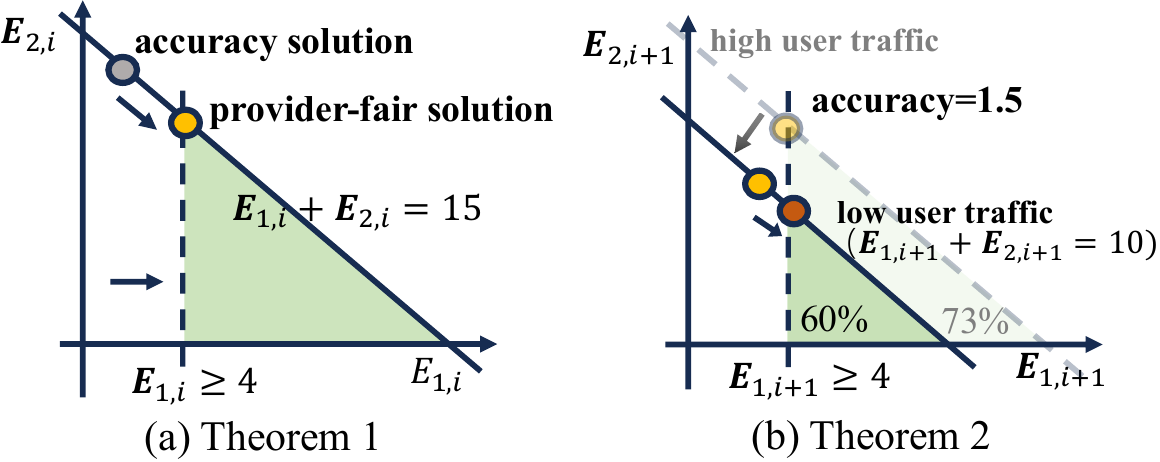}
    \caption{A toy example with two providers to illustrate the optimization process at interval $i$ of Equation~\eqref{eq:unfied_opt}. Provider 1 (x-axis) requires a minimum exposure guarantee of $\bm{M}_{1,i}=4$ and provider 2 (y-axis) has no requirement ($\bm{M}_{2,i}=0$). The green area is the feasible region constructed by the fairness constraint.
(a) Suppose 3 users arrive at interval $i$ and each is recommended 5 items ($3\times 5 = 15$ total exposures). The grey point denotes the accuracy solution $(\bar{\bm{E}}_{1,i},\bar{\bm{E}}_{2,i})$ and the orange point denotes the provider-fair solution $(\bm{E}^{\text{fair}}_{1,i},\bm{E}^{\text{fair}}_{2,i})$, verifying Theorem 1; (b) Suppose the number of users reduced to 2 at interval $i+1$ ($2\times 5 = 10$ total exposures). The red point denotes the provider-fair solution, verifying Theorem 2. 
}    
    \label{fig:toy_example}
\end{figure}

\subsection{Why Provider Fairness Hurts Accuracy?}

From Problem~\eqref{eq:unfied_opt}, we can draw Theorem~\ref{theo:acc_loss_with_cons}:

\begin{theorem}\label{theo:acc_loss_with_cons}
     Let $\bar{f}(\phi,\bm{m})$ be the maximum re-ranking accuracy (i.e., accuracy solution) and $\bar{\bm{E}}_{p,n}$ be the exposure of provider $p$ at interval $n$ without considering provider fairness. When considering fairness in Equation~(\ref{eq:unfied_opt}), if $\bar{\bm{E}}_{p}$ do not satisfy the constraint, i.e., $\sum_{n=1}^N\bar{\bm{E}}_{p,n}< \bm{m}_p$, then $f(\phi,\bm{m})<\bar{f}(\phi,\bm{m})$; otherwise, $f(\phi,\bm{m})=\bar{f}(\phi,\bm{m})$.
\end{theorem}

Proof of Theorem~\ref{theo:acc_loss_with_cons} is given in Appendix~\ref{app:prof_theo1}.
In Theorem~\ref{theo:acc_loss_with_cons}, the accuracy loss brought from provider fairness is through the fairness constraint (i.e., Constraints~\eqref{eq:unfied_opt_con3} from Equation~\eqref{eq:unfied_opt}). If the accuracy-centric solution fails to meet the provider-fair constraint, the compromise will lead to a solution within the feasible region defined by the provider-fair constraint, where accuracy will be compromised.

Figure~\ref{fig:toy_example} (a) presents an illustrative toy example to explain the intuition of Theorem~\ref{theo:acc_loss_with_cons} geometrically. 
From Figure~\ref{fig:toy_example} (a), the accuracy solution is not in the feasible region. Therefore, to satisfy the fairness constraint, the accuracy solution (the grey point) will move to the provider-fair solution (the orange point), resulting in inevitable accuracy loss. 
Meanwhile, the moved distance of two points reflects the degree of accuracy loss~\cite{boyd2004convex}. 

In summary, from the perspective of optimization, provider-fair constraint brings the accuracy loss. 
When the accuracy-maximized solution does not satisfy the fairness constraints, it will inevitably result in a loss of accuracy.
% At the same time, the greater the distance of the solution from the feasible region, the higher the incurred accuracy loss will be.

\subsection{Why Lower Traffic Cause Accuracy Loss?}
Then we will answer this question through Theorem~\ref{theo:low_traffic_cause_harm}

\begin{theorem}\label{theo:low_traffic_cause_harm}
    When the preferences of users arriving at different times are typically random and  independent~\cite{guo2019streaming}, the expectation of accuracy loss $L = \bar{f}(\phi,\bm{m})-f(\phi,\bm{m})$ is negatively correlated to user traffic $\bm{r}_n$ at each interval $n$, i.e., $\mathbb{E}[L] \propto  \frac{1}{\bm{r}_n}$.\end{theorem}

% \begin{remark}
%     Let $S$ be the area of the feasible region constructed by the fairness \xp{constraint, which is proportional to the user traffic $S \propto \prod_{p=1}^{|\mathcal{P}|}(U_n K-M_p)$}. Then the accuracy loss $\bar{a}-f(M_p)$ is proportional to $S$. \xp{delete it?}Meanwhile, the user traffic $U_n$ is proportional to $S$. Therefore, the accuracy loss $\bar{a}-f(M_p)$ is proportional to user traffic $U_n$ at each interval $n$.
% \end{remark}

Proof of Theorem~\ref{theo:acc_loss_with_cons} is given in Appendix~\ref{app:prof_theo2}.
Theorem~\ref{theo:low_traffic_cause_harm} states that lower user traffic will lead to more accuracy loss by affecting the feasible region of the constrained optimization problem.

To better explain Theorem~\ref{theo:low_traffic_cause_harm}, Figure~\ref{fig:toy_example} (b) presents the example of holding the same fairness requirement across interval $i$ and $i+1$ (i.e., $\bm{M}_{1,i}=\bm{M}_{1,i+1}$).
If intervals $i$ and $i+1$ share the same fairness constraints and the traffic at interval $i+1$ drops to two users, the feasible region (green area) 
transits from Figure~\ref{fig:toy_example} (a) to Figure~\ref{fig:toy_example} (b). Consequently, the area of the feasible region reduces from $73\%$ to $60\%$ of the area with no fairness constraints (the area formed by the coordinate axes and the black diagonal line). 
% As a result, the point with accuracy=1.5 (orange point) is no longer in the feasible region and moves to the red point with lower accuracy to satisfy the fairness constraint.
As a result, the provider-fair solution at interval $i$ (represented by the orange point) with an accuracy of $f(\phi,\bm{m}) = 1.5$ no longer complies with the fairness constraint $\bm{E}_{1,i+1}\ge 4$. To meet the constraint, the provider-fair solution at interval $i+1$ must move to the red point, accepting a lower accuracy compared to interval $i$, as indicated in Theorem~\ref{theo:acc_loss_with_cons}.

In summary, the re-ranking accuracy loss stems from the impact of user traffic fluctuations under consistent fairness constraints. As the user traffic varies and fairness constraints remain unchanged, it becomes an important factor that affects the fairness feasible region—lower user traffic narrows the region, while higher user traffic enlarges it. Meanwhile, we also observe that the more narrow the fairness feasible region becomes, the larger the accuracy loss will be.

\begin{algorithm}[t]
        \caption{Online learning of BankFair}
    	\label{alg:bankfair}
    	\begin{algorithmic}[1]
    	\REQUIRE Initial dual variable $\bm{\mu}$, step size $\eta_t$, penalty vector $\bm{\lambda}$, required minimum exposure $\bm{m}_p$, item-provider matrix $\bm{A}$, ranking score $\bm{s}_{u,i}, \forall u,i$.

\ENSURE The re-ranking decision variable $\{\bm{x}_t, t=1,2,...,\infty\}$.
\FOR{$n=1$ to $N$} 
\STATE $// ~~\texttt{Module 1}$.
\STATE $\mathcal{N}=\{n,n+1,\cdots,N \}$.
\STATE  $\hat{\bm{r}}(n)=g\left(\left[\bm{r}_{1}, \bm{r}_{2}, \cdots, \bm{r}_{n-1}\right]\right)$.
\STATE $\bm{D}_p(n)=\alpha K \hat{\bm{r}}(n)$.
\STATE  $\hat{\bm{m}}_{p}(n) = \left[ \hat{\bm{m}}_{p}(n-1) -  \bm{M}_{p,n-1}  + \bm{\beta}_p \right]_{+}, \forall p\in\mathcal{P}$.
\STATE $\hat{\bm{M}}_{p}(n) = \mathrm{TAL}(\mathcal{N},\hat{\bm{m}}_p(n),\bm{D}_p(n)), \forall p\in\mathcal{P}$.
\STATE 
$    \bm{M}_{n}^{\top}  =  \hat{\bm{M}}(n)_1$.
\STATE $// ~~\texttt{Module 2}$.
\STATE Initialize dual variable $\boldsymbol{\mu}=0$.
\STATE Update remaining unearned exposure $\bm{\beta}_n^\top=\bm{M}_{n}^{\top}$.
\FOR{$t=1$ to $\bm{r}_n$}
\STATE  $\bm{x}_{t}=\arg \max _{\bm{x}_t \in \mathcal{X}}\left\{\bm{s}_{u_t}^\top \bm{x}_t /{\hat{\bm{r}}_n}-\bm{\mu}_{t}^{\top} \bm{A}^{\top} \bm{x}_t\right\}$

\STATE $\bm{E}_{n,t}^\top=\arg \max _{\bm{E}_{n,t}\top \leq \bm{\gamma}}\left\{p_{\bm{\lambda}}^*(-\bm{\mu}_t)\right\}$
\STATE Update $\bm{\beta}_n^\top = \bm{\beta}_n^\top - \bm{A}_{t}^{\top}\bm{x}_{t}$.
\STATE 
$\tilde{\bm{g}}_{t}=-\bm{A}_{t}^{\top}\bm{x}_{t}+\bm{E}_{t}.$
\STATE 
$\bm{\mu}_{t+1}=\arg \min _{\bm{\mu} \in \mathcal{D}}\left\langle\tilde{\bm{g}}_{t}, \bm{\mu}\right\rangle+\frac{1}{2 \eta}\left\|\bm{\mu}-\bm{\mu}_{t}\right\|_{w}^{2}.$
\ENDFOR
% \STATE Update $\bm{m}_p = \left[ \bm{D}_p - \bm{M}_{p,n} + \bm{\beta}_{p}\right]_+$.
\ENDFOR
    	\end{algorithmic}
    \end{algorithm}

\section{Our Approach: BankFair}

%BankFair mainly consists of two modules: (1) In the first module, we will get the predicted minimum exposures as the result vectors of the bankruptcy problem. (2) In the second module, we propose an online learning algorithm to obtain the final ranking list $L_K(u)$ for every user $u$ according to the predicted minimum exposure $\bm{M}$.

\begin{figure*}
    \centering
\includegraphics[width=0.95\linewidth]{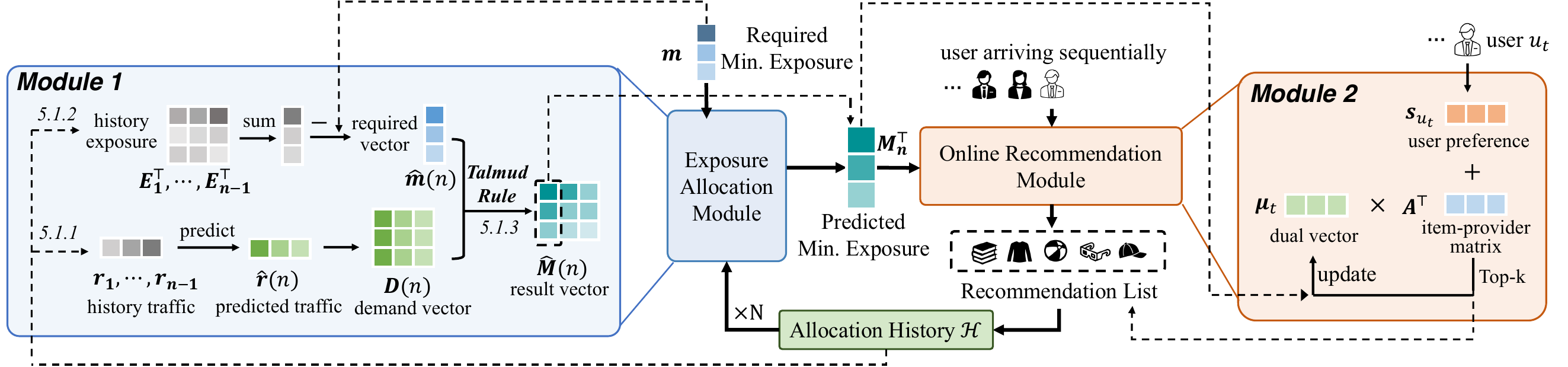}

    \caption{Workflow of the proposed BankFair model}
    \label{fig:overview}

\end{figure*}

In this section, we present our re-ranking algorithm BankFair to guarantee both accuracy and fairness. 
Figure~\ref{fig:overview} depicts the overall workflow of BankFair, and Algorithm~\ref{alg:bankfair} describes the overall procedure. The algorithm consists of the following two modules:
(1) \textit{Line 2-8}: module 1, after a new time interval $n$ starts where we will get the predicted minimum exposures $\bm{M}_n^{\top}$ as the result vectors of the bankruptcy problem. 
(2) \textit{Line 9-18}: module 2, we propose an online learning algorithm to obtain the two-sided re-ranking list $L_K(u_t)$ for every user $u_t$ according to the predicted minimum exposure $\bm{M}_n^{\top}$ at each interval $n$.

\subsection{Module 1: 
Allocating Required Exposures}\label{sec:module1}

In this section, we will get the predicted minimum exposure $\bm{M}_n^{\top}$ at each interval $n$ by transforming the traditional bankruptcy problem into a sequential problem.

In the sequential bankruptcy problem of fair re-ranking, we update the input triplet (upcoming time intervals $\mathcal{N}=\{n,\cdots, N\}$, required minimum exposure $\hat{\bm{m}}_p(n)$, demanded minimum exposure $\bm{D}_p(n)$) and get the result vector $\hat{\bm{M}}(n)$ at each interval $n$.

In this module, we first predict $\bm{D}_p(n)$ in each interval $n$ to be the demand vector using the historical data $\mathcal{H}_{n-1}:=\{\bm{r}_{s},\bm{E}^{\top}_{s}\}_{s=1}^{n-1}$. Then we will use historical data $\mathcal{H}_{n-1}$ to update the remaining required minimum exposure $\hat{\bm{m}}_p(n)$ of the bankruptcy problem.
Finally, we use the input triplet to get the result vector $\hat{\bm{M}}(n)$ and output its first column as the predicted minimum exposure $\bm{M}_n^{\top}$.

\subsubsection{Demand Vector Prediction}
According to Theorem ~\ref{theo:low_traffic_cause_harm}, we know that when the user traffic is low, adopting a larger minimum exposure will result in a greater accuracy loss. Therefore, to guarantee the accuracy level, we need to decrease the fairness degree (i.e., give a low demand for providers) at lower user traffic levels to reduce the accuracy loss. Otherwise, we need to increase the fairness degree (i.e., give a high demand for providers) to compensate for the low user traffic period, fulfilling the exposure requirements $\bm{m}$.
Formally, the demand vector $\bm{D}_{p}(n)\in\mathbb{R}^{|\mathcal{N}|}$ of each provider at interval $n$ can be written as:
\begin{equation}
    \bm{D}_{p}(n) = \alpha K \hat{\bm{r}}(n) =    \alpha K g\left(\left[\bm{r}_{1}, \bm{r}_{2}, \cdots, \bm{r}_{n-1}\right]\right),
\end{equation}
where $\alpha>0$ is a proportional coefficient and $\hat{\bm{r}}(n) = [\hat{\bm{r}}_n,\hat{\bm{r}}_{n+1},\ldots,\hat{\bm{r}}_{N}]$ is the predicted user traffic from interval $n$ to $N$, which is predicted by the time series prediction model $g(\cdot)$ using historical observed data $\mathcal{H}_{n-1}:=\{\bm{r}_{s},\bm{E}^{\top}_{s}\}_{s=1}^{n-1}$. 
% Formally, given the sequence of historical observed data $\mathcal{H}_{n-1}:=\{\bm{r}_{s},\bm{E}^{\top}_{s}\}_{s=1}^{n-1}$, we try to predict the user traffic $\bm{\hat{r}}(n)\in\mathbb{R}^{|\mathcal{N}|}$ in the near future. 
% $
%     \hat{\bm{r}}(n)=g\left(\left[\bm{r}_{1}, \bm{r}_{2}, \cdots, \bm{r}_{n-1}\right]\right),
% $. 
% , which can be replaced with any time series prediction model~\cite{cho2014gru,zhou2021informer}. 
In our experiments, we utilize the well-known time series prediction model Gated Recurrent Unit (GRU)~\cite{cho2014gru}, which can be replaced with any time series prediction model~\cite{donahue2015lstm,zhou2021informer}.

\subsubsection{Update Remaining Minimum Exposure}
Then, we need to update the remaining required minimum exposure $\hat{\bm{m}}_p(n)$ at each interval $n$ as the estate for sequential bankruptcy problem. 
Specifically, at the beginning of each interval $n$, we collect the allocation history $\mathcal{H}_{n-1}:=\{\bm{r}_{s},\bm{E}^{\top}_{s}\}_{s=1}^{n-1}$ to calculate the remaining exposures $\hat{\bm{m}}_{p}(n)\in\mathbb{R}^+$ at interval $n$:
\begin{equation}
    \hat{\bm{m}}_{p}(n) =\left[\hat{\bm{m}}_{p}(n-1) - \bm{E}_{p,n-1} \right]_{+}=\left[\bm{m}_{p} -  \sum_{i=1}^{n-1}\bm{E}_{p,i} \right]_{+}, \forall p\in\mathcal{P},
\end{equation}
where $[\cdot]_{+} \triangleq \max \{0, \cdot\}$.

%First, we define the predicted minimum exposure $\bm{M}_p$ (corresponding to the claim $\bm{d}$ in bankruptcy problem) for each time interval $n, \forall n=1,\cdots,N$.

\subsubsection{Obtain Output Vector}

After getting the input triplet, we will utilize the Talmud rule~\cite{thomson2003axiomatic} in bankruptcy problem to obtain the output vector $\bm{M}_{n}^{\top}$ by getting the first column (i.e., present interval) of the matrix $\hat{\bm{M}}(n)$, i.e.,
$\bm{M}_{n}^{\top}  =  \hat{\bm{M}}_1(n)^{\top}$.
$\hat{\bm{M}}(n)\in\mathbb{R}^{|\mathcal{P}|\times |\mathcal{N}|}$ can be viewed as the predicted future minimum exposure,
which can be calculated by the Talmud rule:
\begin{equation}
\begin{aligned}
\hat{\bm{M}}_{p,i}(n) &= \text{TAL}_i\left(\mathcal{N},\hat{\bm{m}}_p(n),\bm{D}_p(n)\right)\\ & = \begin{cases}
\min\{\frac{\bm{D}_{p,i}(n)}{2},\theta\} & \text{if}~ \hat{\bm{m}}_p(n) \leq \frac{\bm{1}^{\top}\bm{D}_p(n)}{2} \\
\max\{\frac{\bm{D}_{p,i}(n)}{2},\bm{D}_{p,i}(n)-\theta\} & \text{if}~\hat{\bm{m}}_p(n) > \frac{\bm{1}^{\top}\bm{D}_p(n)}{2}
\end{cases},
\end{aligned} 
\label{eq:talmud}
\end{equation}
where $\theta$ is the parameter 
that makes the result vector $\hat{\bm{M}}_{p}(n)$ satisfying 
% the sum of resources allocated to each interval equals the total resource, i.e., 
$\sum_{i\in \mathcal{N}} \hat{\bm{M}}_{p,i}(n)=\hat{\bm{m}}_p(n)$.
% \begin{equation}
% \label{eq:talmud}
% \resizebox{0.92\hsize}{!}{$
% \begin{aligned}
% &\hat{\bm{M}}_p(n) = \text{TAL}(\mathcal{N},\hat{\bm{m}}_p(n),\bm{D}_p(n))\\ & = \begin{cases}
% \text{CEA}\left(\mathcal{N}, \hat{\bm{m}}_p(n), \frac{\bm{D}_p(n)}{2}\right) & \text{if}~ \hat{\bm{m}}_p(n) < \frac{\bm{1}^{\top}\bm{D}_p(n)}{2} \\
% \bm{D}_p(n)-\text{CEA}\left(\mathcal{N}, \bm{1}^{\top}\bm{D}_p(n)-\hat{\bm{m}}_p(n), \frac{\bm{D}_p(n)}{2}\right) & \text{if}~\hat{\bm{m}}_p(n) \geq \frac{\bm{1}^{\top}\bm{D}_p(n)}{2}
% \end{cases},
% \end{aligned} 
% $}
% \end{equation}
% where
% $
%     \text{CEA} (\mathcal{N}, \hat{\bm{m}}_p(n), \bm{D}_p(n))= \bm{b}\in\mathbb{R}^{N},
% $
% and the element $\bm{b}_i= \min \{\bm{D}_{p,i}(n),\lambda^{*}\}$
% with $\lambda^*$ statisfies 
% $
% \sum_{i=1}^N \min \{ \bm{D}_{p,i}(n),\lambda^* \} = \hat{\bm{m}}_p.
% $

Through the Talmud rule $\text{TAL}(\cdot)$, we can allocate the required minimum exposure $\hat{\bm{m}}_p(n)$ into each interval according to the bankruptcy problem. 
The Talmud rule can be seen from two conditions: (1) If the minimum exposure $\hat{\bm{m}}_p(n)$ does not exceed half of the total demands $\bm{1}^{\top}\bm{D}_p(n)/2$: $\hat{\bm{m}}_p(n)$ will be distributed equally among each time interval $n$ until $\hat{\bm{M}}_{p,n}(n)$ meets half of the demand $\bm{D}_{p,n}(n)/2$.
(2) If the minimum exposure $\hat{\bm{m}}_p(n)$ exceed the total demands $\bm{1}^{\top}\bm{D}_p(n)/2$: $\hat{\bm{m}}_p(n)$ will be allocated to equalize the difference between each allocation $\hat{\bm{M}}_{p,n}(n)$ and its demand $\bm{D}_{p,n}(n)$.

\subsection{Module 2: Online Recommendation}
%In this section, we formulate our provider-fair re-ranking task as a fluctuating resource allocation problem~\cite{katoh1998resourceallocationproblems}. 
After getting the predicted minimum exposure $\bm{M}_n^{\top}$ as the provider fairness constraint at each interval $n$, we need to get the 
recommendation list $L_K(u_t)$ for each user $u_t$.

Following ~\cite{xu2023p,surer2018multistakeholder}, we first use an integer-programming (IP)-based method to formulate the two-sided re-ranking problem in an offline scenario. Then, we adapt the offline algorithm into an online version to adapt the sequential arriving characteristic of users.

%propose online learning algorithm to solve it efficiently in the online scenario.

%quad \forall p \in \mathcal{P}\\
\subsubsection{BankFair for Offline Scenario}
At each interval $n$, we formulate the two-sided re-ranking task as a resource allocation problem~\cite{balseiro2021regularized}:
\begin{equation}\label{eq:offline}
\begin{aligned}
\max _{x\in \mathcal{X}} \quad \frac{1}{\bm{r}_n}\sum_{u_t \in \mathcal{U}_n}\bm{s}_{u_t}^\top \bm{x}_t\quad 
\text { s.t. }  \quad \bm{E}_{n}^{\top} = \sum_{t} \bm{A}^{\top} \bm{x}_{t} \geq \bm{M}_{n}^{\top}, \\
%& \bm{E}_{n}^{\top} = \sum_{t=1}^{\bm{r}_n} \bm{A}^{\top} \bm{x}_{t},
\end{aligned}
\end{equation}
where $\bm{x}_t \in \mathcal{X}$ is the decision variable that decides whether item $i$ is recommended to user $u_t$. Given ranking list length $K$, 
$\mathcal{X} =\{ \bm{x}_t 
\in \{0,1\}^{|\mathcal{I}|} |  \bm{1}^{\top}\bm{x}_{t}=K \}$. Specifically, for each item $i$, $\bm{x}_{t,i} = 1$ if it is added to the re-ranking list $L_K(u_t)$ with size $K$, otherwise $\bm{x}_{t,i} = 0$. 
$\bm{A}$ is the item-provider adjacent matrix, where $\bm{A}_{i,p}=1$ indicates item $i \in \mathcal{I}_{p}$, and 0 otherwise.

In Equation~(\ref{eq:offline}), the objective is to maximize the accuracy while ensuring the fairness constraint. The accuracy $a(n;\bm{x})$ denotes the sum of the ranking scores $\bm{s}_{u_t}$ (see Section~\ref{sec:formulation}) for each re-ranking list $L_K(u_t)$ on average. The fairness constraint means that provider $p$ must receive at least $\bm{M}_{p,n}$ exposures of items corresponding to $p$ (the item-provider correspondence relation can be used $\bm{A}$ to represent) at interval $n$. Since we have obtain the traffic-adaptive minimum exposure $\bm{M}_n$ in the previous section, we are not imposing any minimum constraints on accuracy.

\subsubsection{BankFair for Online Scenario.}
In real online recommendation scenarios, users usually arrive one after the other sequentially (see Figure~\ref{fig:overview}) and we should give user $u_t$ recommendation list $L_K(u_t)$ immediately.
Hence, we develop an online version of the offline algorithm at each interval $n$.

To efficiently solve Equation~\eqref{eq:offline}, we utilize Lagrangian relaxation, transforming Equation~\eqref{eq:offline} into a simpler regularized optimization problem utilizing the following Theorem.

\begin{theorem}\label{theo:Solve_BankFair}
   The Equation~(\ref{eq:dual}) is a relaxed dual problem of Equation~\eqref{eq:offline}:
    \begin{equation}\label{eq:dual}
 W_{Dual} =  \min _{\bm{\mu} \in \mathcal{D}} \left[h^{*}\left(\bm{A} \bm{\mu} \right)\right]+p_{\bm{\lambda}}^{*}(-\bm{\mu}),
    \end{equation}
    where $\mathcal{D}=\left\{\bm{\mu} \in \mathbb{R}^{|\mathcal{P}|} \mid \bm{\mu} \geq -\bm{\lambda}\right\}$ is the feasible region of dual variable $\bm{\mu}$ and 
    \[h^*(\bm{A} \bm{\mu}) = \max _{\bm{x}_{t} \in \mathcal{X}}\sum_{t=1}^{\bm{r}_n}\left[\bm{s}_{u_t}^\top\bm{x}_{t} /\bm{r}_n -\bm{\mu}^{\top}  \bm{A}^{\top} \bm{x}_{t}\right],
    \] 

    \[
    p_{\bm{\lambda}}^*(-\bm{\mu})=\max _{\bm{E}_n\top \leq \bm{\gamma}}\left[ -\sum_{i=1}^{|\mathcal{P}|}  \bm{\lambda}_i \left[\bm{M}_{n,i}^\top - \bm{E}_{n,i}^\top\right]_+ + \bm{\mu}^{\top} \bm{E}_n^\top \right].
    \]
    % \[
    %     p_{\bm{\lambda}}(\bm{E}_n^\top)=-\sum_{i=1}^{|\mathcal{P}|}  \bm{\lambda}_i \left[\bm{M}_{n,i}^\top - \bm{E}_{n,i}^\top\right]_+.
    % \]
\end{theorem}

% The proof of the Theorem can be seen in Appendix~\ref{theo:proof_theo3}.

\begin{lemma}
\label{lemma}
The conjugate function $p_{\bm{\lambda}}^*(\cdot)$ has a close form $p_{\bm{\lambda}}^*(-\bm{\mu})=\bm{\mu}^{\top} \bm{M}_{n}^\top+\sum_{p=1}^{|\mathcal{P}|}\left(\bm{\gamma}_{p}-\bm{M}_{p,n}\right) \max \left(\bm{\mu}_{p}, 0\right)$, $\forall \bm{\mu} \in \mathcal{D}$. The optimal dual variable is: $\bm{E}_{p,n}^*(-\bm{\mu})=\bm{M}_{p,n}$ if $\bm{\mu}_p \in\left[-\bm{M}_{p,n}, 0\right)$ and $\bm{E}_{p,n}^*(-\bm{\mu})=\bm{\gamma}_{p}$ if $\bm{\mu}_p \geq 0$.
\end{lemma}

Due to space limitation, we put the proof of Theorem~\ref{theo:Solve_BankFair} and Lemma~\ref{lemma} in the github repository\footnote{\url{https://github.com/shawnye2000/BankFair}\label{fn:github}}.

Within each time interval $n$, the algorithm computes the decision variable $\bm{x}_t$ based on $h^*(\bm{A} \bm{\mu})$ after a new user $u_t$ arrives. Then, the algorithm updates the auxiliary variable $\bm{E}$. Then, we obtain the stochastic subgradient of the dual function and use the weighted, projected subgradient method to update the dual variable $\boldsymbol{\mu}_t$.

\begin{figure*}
    \centering
\includegraphics[width=1\linewidth]{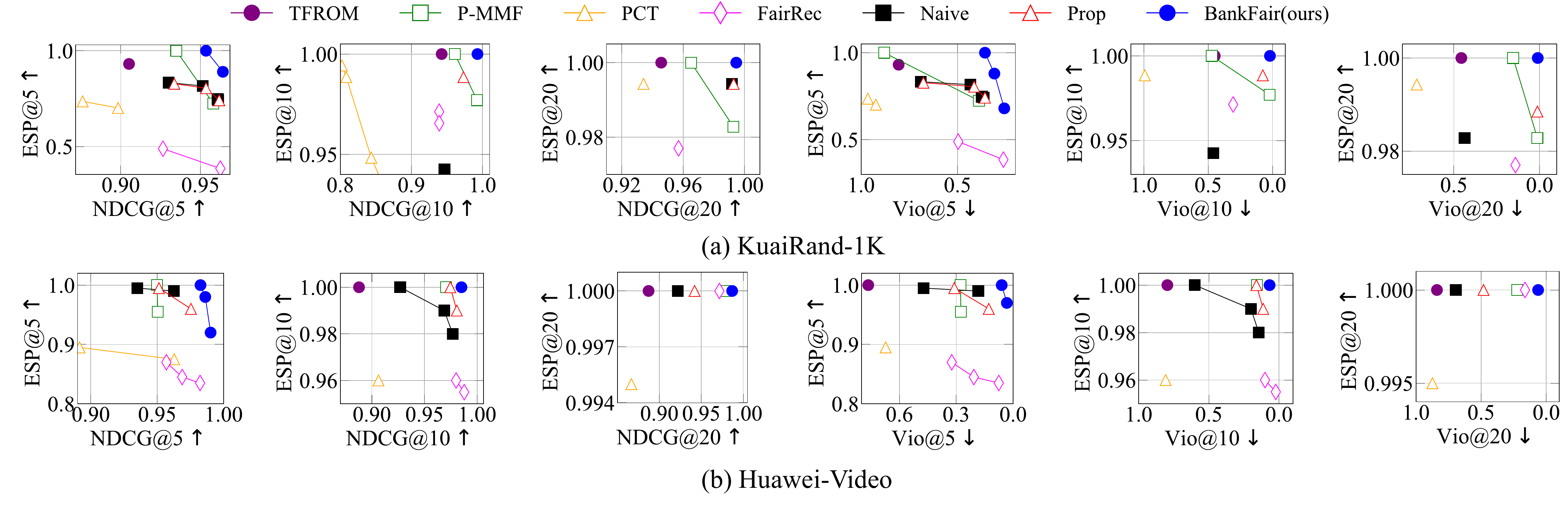}
    \caption{Pareto frontier of two different datasets with different top-$K$ ranking. Y-axis shows ESP@K metric, while X-axis shows NDCG@K metric and Vio@K metric. $\uparrow$ means higher values are better and $\downarrow$ favors lower values. According to the size of the dataset, for KuaiRand-1K, we set $\bm{m}_p = 1000$ and for Huawei-Video, we set $\bm{m}_p = 100$. For both datasets, we set $\phi=0.95$.
    }
    \label{fig:pareto}
\end{figure*}

\section{Experiment}
In this section, we conduct experiments on two real datasets. The source codes and experiments have been shared at  GitHub\footref{fn:github}.

\subsection{Experimental Settings}

\subsubsection{Datasets}
To accurately represent real user traffic, we conducted experiments on two real-world recommendation datasets. The interaction data in both datasets were randomly sampled from actual user traffic in short video applications.

\textbf{KuaiRand-1K}\footnote{\url{https://kuairand.com/}}~\cite{gao2022kuairand}: a dataset collected from the video-sharing mobile app, \textit{KuaiShou}. The data are all derived from interactions from 8th April 2022 to 8th May 2022 from \textit{KuaiShou}.
We use the standard part, which contains randomly sampled user traffic of 302870 interactions from 933 users on 6825 videos with 174 providers.

\textbf{Huawei-Video}: a dataset collected from the short video platform in \textit{Huawei Browser} from 2 Jan. 2024 to 8 Jan. 2024, spanning a week. It contains 19355 users, 5364 items, and 200 providers, totaling 118765 interactions. 

For KuaiRand-1K, we use the interactions before 23rd April for training and 23rd April to 8th May for testing. For Huawei-Video, we use interactions before 5th Jan. for training and 6th Jan. to 8th Jan. for testing.

\subsubsection{Algorithm Update Setting}
We divide the datasets into a training set and a testing set chronologically.
 The testing set is further divided into $N$ update intervals.
To simulate the algorithm update process in real scenarios, we update the base model (e.g., LightGCN~\cite{he2020lightgcn}) at the beginning of each interval using all the historical interactions accumulated.

\subsubsection{Evaluation}
\label{sec:evaluation} For accuracy metrics, 
following previous practices in~\cite{wu2021tfrom,xu2023p,biega2018equity}, we use the normalized discounted cumulative gain (NDCG) to measure the accuracy of recommendation, which is defined as the discounted cumulative relevance score between the re-ranking list $L_K(u_t)$ and the user's original recommendation $L_K^{\text{ori}}(u_t)$.  The average NDCG for all users can be defined as:

% \begin{equation}
%    \mathrm{NDCG}@K=\frac{1}{N} \sum_{n=1}^N \mathrm{NDCG}_n@K,
%    %
% \end{equation}
% where
\begin{equation}
\label{eq:ndcg}
    \mathrm{NDCG} @ \mathrm{K}=\frac{1}{\sum_{j=1}^N |\mathcal{U}_j|}
    \sum_{n=1}^{N} \sum_{u_t\in\mathcal{U}_n} \mathrm{NDCG}_{u_t}@\mathrm{K},
\end{equation}
where $\mathrm{NDCG}_{u_t}@\mathrm{K} = \frac{\sum_{i \in L_{K}\left(u_{t}\right)} \bm{s}_{u_{t}, i} / \log \left(\operatorname{rank}_{i}+1\right)}{\sum_{i \in L_{K}^{\text{ori}}\left(u_{t}\right)} \bm{s}_{u_{t}, i} / \log \left(\operatorname{rank}_{i}^{\text{ori}}+1\right)}$, 
$\text{rank}_i$ and $\text{rank}_i^{\text{ori}}$ are the ranking positions of the item $i$ in $L_K(u_t)$ and $L_K^{\text{ori}}(u_t)$.  

To evaluate the extent to which a model sacrifices user experience to achieve fairness, we use the proportion of users receiving
accuracy below the minimum requirement $\phi$, which is defined as the accuracy violation (Vio):
\begin{equation}
    \mathrm{Vio} @ \mathrm{K}=\frac{1}{\sum_{j=1}^N |\mathcal{U}_j|}\sum_{n=1}^{N} \sum_{u_t\in\mathcal{U}_n} \mathbb{I}\left( \mathrm{NDCG}_{u_t}@\mathrm{K}< \phi \right),
\end{equation}
% \begin{equation}
%     \mathrm{Vio} @ \mathrm{K}=\frac{1}{\sum_{j=1}^N |\mathcal{U}_j|}\sum_{n=1}^{N} \sum_{u_t\in\mathcal{U}_n} \max\{0, \phi -   \mathrm{NDCG}_{u_t}@\mathrm{K} \},
% \end{equation}
where $\mathbb{I}(\cdot)=1$ if the condition $(\cdot)$ is satisfied otherwise 0. If all users satisfies the 
 minimum accuracy requirement, then $\mathrm{Vio}@K$ becomes 0.

For fairness evaluation, we measure the provider group's satisfaction with their exposure over a period.
Following~\cite{wang2023uncertainty,patro2020fairrec, xu2024fairsync}, we use the metric of enough satisfaction group (ESP):
\begin{equation}
\mathrm{ESP@K}=\frac{1}{|\mathcal{P}|} \sum_{p \in \mathcal{P}} \mathbb{I}\left(\left[\sum_{n=1}^{N}\sum_{u_t\in\mathcal{U}_n} \sum_{i \in L_{K}\left(u_{t}\right)} \mathbb{I}\left(i \in \mathcal{I}_{p}\right)\right]\geq \bm{m}_{p}\right).
\end{equation}

 The value of ESP ranges between 0 and 1. When all the providers get exposures no less than the minimum exposure guarantee $\bm{m}_p$, then ESP becomes 1.

\subsubsection{Baselines}
We compare with the following two-sided fairness-aware baselines:
\textbf{P-MMF}~\cite{xu2023p}: An integer-programming (IP)-based re-ranking method ensures the exposure of worst-off provider; \textbf{FairRec}~\cite{patro2020fairrec}: A method guarantees the minimum exposure for providers and employs a greedy strategy to uphold user fairness; \textbf{TFROM}~\cite{wu2021tfrom}: A two-sided re-ranking method which improves provider exposure and amortizes the accuracy loss among users to guarantee accuracy; \textbf{PCT}~\cite{wang2023pct}: A max-margin-relevance-based method to ensures a target item exposure distribution and reduce accuracy loss for users.

We also consider two baselines that take a simple allocation rule in Module 1 (See Section~\ref{sec:module1}) while maintaining Module 2 as same as ours.
\textbf{Naive}: This algorithm sets $\bm{M}_{p,n}=\bm{m}_p/2$ if the predicted user traffic $\hat{\bm{r}}_n$ exceeds threshold $c$ and $\bm{M}_{p,n}=0$ otherwise. In our experiment, we set $c$ as the mean value of the upcoming predicted user traffic $c=\mathrm{Mean}(\{\hat{\bm{r}}_i\}_{i=n}^N)$; 
\textbf{Prop}: This algorithm determines the minimum exposure $\bm{M}_{p,n}$ based on the proportion of predicted traffic for the current interval $n$ to the overall upcoming predicted traffic: $\bm{M}_{p,n}=\frac{\hat{\bm{r}}_{n}}{\sum_{i=n}^N \hat{\bm{r}}_{i}} \bm{m}_p$.

\subsubsection{Implementation Details.}
For the BankFair models, the proportionality coefficient $\alpha$ was set to $\alpha = \frac{k\sum_{p=1}^{|\mathcal{P}|}\bm{m}_p}{|\mathcal{P}|\sum_{n=1}^{N} K \bm{r}_n }$ and $k$ was tuned among $[1,2]$. Meanwhile, the bankruptcy problem has a requirement: the demand resource should exceed the estate, therefore, $\sum_{n=1}^N \alpha K \hat{\bm{r}}_n \ge \bm{m}_p, \forall p\in\mathcal{P}$.
For the violation penalty vector $\bm{\lambda}$, we set $\bm{\lambda}_p= \beta \frac{\max_p(|\mathcal{I}_p|)}{|\mathcal{I}_p|} + \frac{1-\beta}{|\mathcal{P}|}$, where $\beta\in [0, 1]$ is a factor controlling the importance on small providers (providers with lower $|\mathcal{I}_p|$).
Following the practices in~\cite{wu2021tfrom,xu2023p}, we set $\bm{\gamma}$ based on the number of items provided by the providers: $\bm{\gamma}_p =K \hat{\bm{r}}_n\left|\mathcal{I}_{p}\right| /|\mathcal{I}|, \forall p\in\mathcal{P}$.

\begin{figure}
    \centering
\includegraphics[width=1\linewidth]{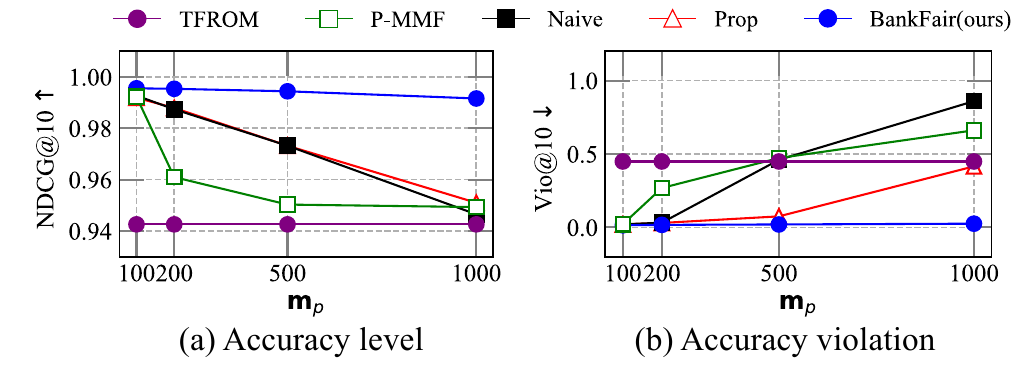}
\caption{\mbox{NDCG@10 and Vio@10 w.r.t. minimum exposure $\bm{m}_p$.}}
    \label{fig:ablation_study1}
\end{figure}

\begin{figure}
    \centering
\includegraphics[width=1\linewidth]{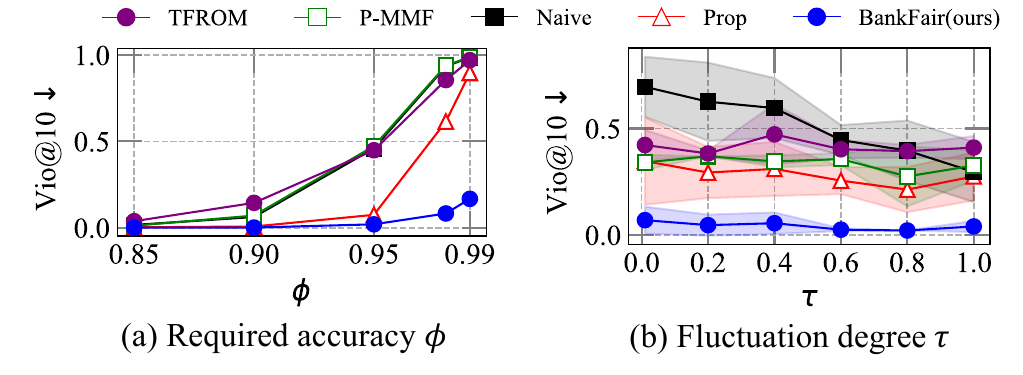}
    \caption{(a) Vio@10 under different required minimum accuracy $\phi$. (b) The effect of the fluctuation degree $\tau$ on Vio@10.}
    \label{fig:softmax}
\end{figure}

\subsection{Experimental Results}
\label{sec:exp_result}

In this section, we conduct experiments to test the performance of BankFair and baselines in terms of short-term accuracy and long-term fairness, on KuaiRand-1K and Huawei-Video.
In all the experiments, LightGCN~\cite{he2020lightgcn} is chosen as the base model to generate the relevance score $\bm{s}$. We set the algorithm update interval to $24$ hours (1 day), which is the most commonly used setting. Considering different dataset sizes, we set $\bm{m}_p=1000$ on KuaiRand-1K and $\bm{m}_p=100$ on Huawei-Video. We set $\phi=0.95$ for calculating the Vio@K metric for both datasets.
% To make fair comparisons, we select the result with the best robustness performance from those achieving the highest ESP metric for all baselines.

Figure~\ref{fig:pareto} shows the Pareto frontiers ~\cite{lotov2008pareto} of ESP@K-NDCG@K and ESP@K-Vio@K on two datasets with different ranking sizes $K$.
Note that, for the hyperparameter tuning for each model, we only retain the points on the Pareto front, therefore, in Figure~\ref{fig:pareto}, some methods will have only one point, while others have multiple.

% The Pareto frontiers are drawn by tuning different parameters of the models and choosing the (NDCG@K, ESP@K) and (Vio@K, ESP@K) points with the best performances.
From the Pareto frontiers, we find that BankFair Pareto dominates the baselines (i.e., the BankFair points are at the upper right corner), which means BankFair achieves higher accuracy level (NDCG@K) and guarantee lower short-term accuracy violation (Vio@K) under the same fairness (ESP@K) level. Furthermore, we find the proposed BankFair can satisfy the 
required minimum exposure for all providers (i.e., ESP=100\%) and ensure the minimum accuracy for nearly all users (i.e., Vio$\approx$0\%) on all datasets under all top-K settings.

Meanwhile, we notice that when ranking list size $K$ is small (e.g., $K=5$), most baselines fail to meet the short-term accuracy requirement for the majority of users (i.e., maintaining a high Vio@K) while achieving 100\% ESP@K. However,  BankFair guarantees the minimum accuracy requirements for nearly all users (Vio@K$\approx$0\%) and satisfies the required minimum exposure for all providers (i.e., ESP@K=100\%) simultaneously. 

All the experimental results verify that BankFair splendidly guarantees the short-term accuracy requirements and the long-term required minimum exposure of all providers.

% \begin{figure}
%     \centering
% \includegraphics[width=1\linewidth]
% {fig/interval_len.pdf}
% \caption{NDCG@10 and ESP@10 w.r.t. gap time $h$.}
%     \label{fig:interval}
% \end{figure}
\subsection{Experimental Analysis}
We conducted experiments to analyze BankFair on KuaiRand-1K. In this section, we compare the best-performed baselines in Section~\ref{sec:exp_result}. 
% For consistency, the marker of models in the figures in this section is identical to those in Figure~\ref{fig:pareto}."

\subsubsection{\textbf{Influence of Required Minimum Exposure} $\bm{m}$.}

%In the main experiment (Table~\ref{tab:main_exp}), we only test the $\bm{m}=100$. However, 
In real applications, the platforms often change the required minimum exposure $\bm{m}$ to fit different applications (e.g., development stage, incentive policy)~\cite{bardhan2022more}. 
In this experiment, we investigate the performance of BankFair as the required minimum exposure varies $\bm{m}_p \in [100, 1000], \forall p\in\mathcal{P}$. 
% In the analysis, we only consider the baselines that satisfy the minimum exposure guarantees of all providers (i.e., those achieved $100\%$ ESP in Figure~\ref{fig:pareto}).
To make fair comparisons for all selected models, we select the result with the best NDCG@K and Vio@K performance with at least $95\%$ ESP@K under different $\bm{m}_p$.

% Figure~\ref{fig:ablation_study1} illustrates the results where the x-axis denotes the value of $\bm{m}_p$, and y-axis denotes NDCG@10 and Vio@10 in Sub-figure~\ref{fig:ablation_study1} (a) and (b), respectively.
From Vio@10 and NDCG@10 curves in Figure~\ref{fig:ablation_study1}, we can see that the proposed BankFair achieves better accuracy (higher NDCG@10 and lower Vio@10) in comparison to the baselines under different required minimum exposure $\bm{m}_p$ levels. 
Moreover, the accuracy metric of most baselines (except TFROM, as there are no hyperparameters to adjust) worsens by a large margin (i.e., NDCG@K decreases, Vio@K increases) when the fairness requirement becomes strict (i.e., $\bm{m}_p$ increases). However, BankFair can maintain relatively stable and high accuracy under different required minimum exposure $\bm{m}_p$ levels. 
% Similar phenomena are also observed at different $K$ values (i.e., $K=5,20$, and the results are reported in Appendix~\ref{app:different_mp}).

\subsubsection{\textbf{Influence of Required Minimum Accuracy} $\phi$.}
We conduct experiments to validate the effectiveness of BankFair under different required minimum accuracy $\phi$. Sub-figure~\ref{fig:softmax} (a) shows the accuracy violation (Vio@K) performance of BankFair and the best-performeed baselines under different required minimum accuracy $\phi\in [0.85,0.99]$. From Sub-figure~\ref{fig:softmax} (a), we observe that BankFair guarantees a very low Vio@K under any required accuracy level and greatly outperforms all selected baselines. 
It is worth noting that BankFair maintains low Vio levels even at higher values of $\phi$ (i.e., more strict accuracy requirement).

\subsubsection{\textbf{Influence of User Traffic Fluctuation Degree} $\tau$}
In real-world scenarios, user traffic fluctuations can vary significantly, ranging from intense to relatively steady~\cite{beel2017MrDLib,pires2015youtube}.
In this experiment, we  validate the effectiveness of our method under varying degrees of traffic fluctuations. To realize different fluctuation degree, we set a temperature coefficient $\tau$  with which we re-sample the user traffic $\bm{r}_n$ at interval $n$ with the probability $p_n$, where $p_n=\text{Softmax}([\bm{r}_1,\bm{r}_2,\cdots, \bm{r}_n]/\tau)$.
Intuitively, the user traffic disparities become substantial when $\tau \rightarrow 0$, and the user traffic tends to become nearly uniform when $\tau \rightarrow 1$.

Sub-figure~\ref{fig:softmax} (b) shows the Vio@10 under varying fluctuation degree $\tau$. The shaded area indicates the 95\% confidence intervals of $t$-distribution under 5 experiments with different random seeds.
From Sub-figure~\ref{fig:softmax} (b), we observe that BankFair can achieve lower Vio@K under different degrees of fluctuation, especially when the fluctuation is intense (i.e., $\tau \in [0.0,0.2]$). 
Meanwhile, We can observe that as $\tau$ decreases (i.e., more intense fluctuations), Vio@K of most baselines, especially Naive, increases greatly, which supports our theoretical analysis in Section~\ref{sec:harm} that lower traffic strengthens accuracy loss.

\begin{table}[t]
    \small
    \centering
\caption{Performances of BankFair and baselines by using BPR and NCF as the base models. The experimental settings are identical to that reported in Section~\ref{sec:exp_result}. $*$ means the improvements over the baseline that achieve ESP=100\% are statistically significant ($t$-tests and $p$-value < 0.05). }\label{tab:base_model}
    \scalebox{0.88}{
% \resizebox{1\columnwidth}{15mm}{
    \begin{tabular}{l  ccc ccc}
    \toprule
  Base model & \multicolumn{3}{c}{\textbf{BPR}} &\multicolumn{3}{c}{\textbf{NCF}}\\ 
   \cmidrule(r){2-4} \cmidrule(r){5-7}  
     Method &  NDCG $\uparrow$ & Vio $\downarrow$  & ESP $\uparrow$ &  NDCG $\uparrow$ & Vio $\downarrow$  & ESP $\uparrow$ \\ \midrule
  FairRec   & 0.9561   & \underline{0.1746} & 0.8333 &  0.9429   & 0.2408  &  0.9598     \\ 
 PCT   & 0.7879  & 0.9957  &  0.9828    & 0.7773  & 0.9957 & 0.9828\\ 
          TFROM & 0.9158  & 0.7142  &  1.0000 & 0.9430  & 0.4589  &  1.0000 \\ 
          P-MMF& 0.9403   & 0.5679      & 1.0000   & 0.9571  & 0.3009  &  1.0000    \\ 
         \cline{1-7}
          Naive    & 0.9377  &  0.6356  &  0.9310  & 0.9458 &  0.4926  & 0.9713    \\
          Prop   & \underline{0.9654}  & 0.1850  &  0.9540   & \underline{0.9758}  &  \underline{0.0590}  & 1.0000 \\
       \textbf{BankFair(ours)}  & \textbf{0.9806}$^*$ & \textbf{0.1179}$^*$   &  1.0000 &    \textbf{0.9957}$^*$  & \textbf{0.0024}$^*$  & 1.0000  \\
       \hline
   \textbf{Improv.}  & \textbf{1.6\%} $\uparrow$ & \textbf{36.3\% }$\downarrow$  &   &  \textbf{2.0\%}$\uparrow$  & \textbf{95.9\%}$\downarrow$  & \\
        \bottomrule
    \end{tabular}
    }
\end{table}

\subsubsection{\textbf{Ablation Study on Different Base Models}.}
 We conduct experiments to test the performances of BankFair with the two widely used ranking models of \textbf{BPR}~\cite{rendle2012bpr} and \textbf{NCF}~\cite{he2017neuralmf} as the base ranking model to generate the relevance score $\bm{s}$. We select the result with the best NDCG@K and Vio@K performance with highest ESP@K.
As shown in Table~\ref{tab:base_model}, we observe that BankFair achieves 100\% ESP@K, better accuracy, and lower accuracy violation than baselines under different base models.
% Similar phenomena can also be observed under different $K$ values (See Appendix~\ref{app:base_model}).

\section{Conclusion and Future works}
This paper emphasizes the significance of guaranteeing both short-term accuracy and long-term fairness under fluctuating user traffic in two-sided recommendation platforms. By formalizing the two-sided re-ranking as a constrained optimization problem, we theoretically analyze the root cause of accuracy loss under fluctuating user traffic. Inspired by the analysis and the Talmud rule in the bankruptcy problem as the exposure allocation principle, we propose a new re-ranking model called BankFair to balance accuracy-fairness well under fluctuating user traffic applications.
Experiments on two real-world recommendation datasets show that BankFair can outperform all baselines in terms of both accuracy and fairness. 
% Experiments on a public and a real industrial dataset show that BankFair can outperform all baselines in terms of both accuracy and fairness.
In future work, we will also explore other forms of fairness functions, such as user fairness, under various user traffic scenarios.

%\xp{In this paper, we only considered provider fairness. However, user fairness is also an important factor in two-sided platforms that cannot be ignored. In the future, we will take it into account.}

%%

\begin{acks}
  This work was funded by the National Key R\&D Program of China (2023YFA1008704), the National Natural Science Foundation of China (No. 62377044), Beijing Key Laboratory of Big Data Management and Analysis Methods, Major Innovation \& Planning Interdisciplinary Platform for the ``Double-First Class'' Initiative, funds for building world-class universities (disciplines) of Renmin University of China, and PCC@RUC.  
\end{acks}

\appendix
\section{Appendix}

\subsection{Proof of Theorem~\ref{theo:acc_loss_with_cons}}
\label{app:prof_theo1}
\begin{proof}
We first give the formal definition of $\bar{f}(\phi,\bm{m})$, i.e., $$\bar{f}(\phi,\bm{m}) = \max \sum_{n=1}^N a(n;\bm{E}) \quad \text{s.t.} \quad \text{Constraint~\eqref{eq:unfied_opt_con1}}.$$

It's important to note that Constraint~\eqref{eq:unfied_opt_con1} will be naturally satisfied without Constraint~\eqref{eq:unfied_opt_con3} in the problem $\bar{f}(\phi,\bm{m})$. This is because, according to the re-ranking accuracy definition (see Section~\ref{sec:evaluation}), the re-ranking accuracy $a(n;\bm{E})$ will be 1 without the fairness constraint, satisfying $\phi$ naturally.

We know that the difference between problem $\bar{f}(\phi,\bm{m})$ and problem $f(\phi,\bm{m})$ is Constraint~\eqref{eq:unfied_opt_con3}. 
Therefore, we can conclude that $\bar{f}(\phi,\bm{m})\geq f(\phi,\bm{m})$ always holds, because the feasible region of $\bar{f}(\phi,\bm{m})$ contains the feasible region of $f(\phi,\bm{m})$. 

We prove Theorem~\ref{theo:acc_loss_with_cons} in two condition:

(1) If $\bar{f}(\phi, \bm{m})$ satisfies Constraint~\eqref{eq:unfied_opt_con3}, then it falls in the feasible region of $\bar{f}(\phi,\bm{m})$. 
Also, since it is the solution which maximize the Objective~\eqref{eq:unfied_opt}, $\bar{f}(\phi, \bm{m})=\bar{f}(\phi,\bm{m})$;

(2) If $\bar{f}(\phi, \bm{m})$ does not satisfy Constraint~\eqref{eq:unfied_opt_con3}, $\bar{f}(\phi, \bm{m})\neq \bar{f}(\phi,\bm{m})$. Since $\bar{f}(\phi,\bm{m})\geq f(\phi,\bm{m})$, then $\bar{f}(\bm{m}) > f(\bm{m})$.

\end{proof}

\subsection{Proof of Theorem~\ref{theo:low_traffic_cause_harm}}
\label{app:prof_theo2}
\begin{proof}
Let $S_{\text{fair}}$ and $S$ denote the areas of feasible regions of $f(\phi,\bm{m})$ and $\bar{f}(\phi,\bm{m})$), respectively.
First, we can know that $S \geq S_{\text{fair}}$ since the feasible region without Constraint~\eqref{eq:unfied_opt_con3} indeed contains the feasible region with Constraint~\eqref{eq:unfied_opt_con3}.
Then, we can compute $S_{\text{fair}}/S$:
\begin{equation*}
\small
\begin{aligned}
 \frac{S_{\text{fair}}}{S} &= \prod_{l=1}^{|\mathcal{P}|}\frac{\bm{r}_nK-\sum_{p=1}^{l}\bm{M}_{p,n}}{\bm{r}_nK-\sum_{p=1}^{l-1}\bm{M}_{p,n}}= \frac{\bm{r}_nK-\sum_{p=1}^{|\mathcal{P}|}\bm{M}_{p,n}}{\bm{r}_nK}=  1 - \frac{\sum_{p=1}^{|\mathcal{P}|}\bm{M}_{p,n}}{\bm{r}_nK}.
\end{aligned}
\end{equation*}

% From the equation above, we can see that the ratio of two feasible region areas $1- \frac{S_{\text{fair}}}{S}\propto \frac{1}{\bm{r}_n}$.

According to Theorem~\ref{theo:acc_loss_with_cons}, we know $\bar{f}(\phi,\bm{m}) \geq f(\phi,\bm{m}) > 0$ always holds.
Let $f(\phi,\bm{m})=k \bar{f}(\phi,\bm{m})$ and $\tau = (1-k)\bar{f}(\phi,\bm{m})$, where $k\in (0,1]$. 
Due to the randomness of the user's preference, we can write the expectation of accuracy loss $\mathbb{E}[L]$:
\begin{equation*}
    \begin{aligned}
        \mathbb{E}[L] = &\mathbb{E}[\bar{f}(\phi,\bm{m}) - f(\phi,\bm{m})] =\mathbb{E}[\tau] \\
        = &P\left(k\neq 1\right) \mathbb{E}[\tau|k\neq 1]+  P\left(k= 1\right) \mathbb{E}[\tau|k= 1]\\
     = &P\left(k\neq 1\right) \mathbb{E}[\tau|k\neq 1] 
     =  P\left(\bar{f}(\phi,\bm{m}) > f(\phi,\bm{m})\right) \mathbb{E}[\tau|k\neq 1] .
    \end{aligned}
\end{equation*}
Here, we use the Monte Carlo method~\cite{rubinstein2016montecarlo} to approximate \\
$P\left(\bar{f}(\phi,\bm{m}) > f(\phi,\bm{m})\right)\approx 1-\frac{S_{\text{fair}}}{S}$. Then we can have:\\$\mathbb{E}[L] \approx  \left( 1-\frac{S_{\text{fair}}}{S} \right) \mathbb{E}[\tau|k\neq 1]$,
where $\mathbb{E}[\tau|k\neq 1]$ is irrelevant to the user traffic $\bm{r}_n$. Its exact expression is out of the scope of this paper.
Therefore, we can conclude that the expectation of accuracy loss $L$ is proportional to $\frac{1}{\bm{r}_n}$, i.e., $\mathbb{E}[L] \propto 1-\frac{S_{\text{fair}}}{S} \propto \frac{1}{\bm{r}_n}$.
\end{proof}

\newpage
\bibliographystyle{ACM-Reference-Format}
\balance
\bibliography{sample-base}


\begin{thebibliography}{57}

%%% the Web address.

\ifx \showCODEN    \undefined \def \showCODEN     #1{\unskip}     \fi
\ifx \showDOI      \undefined \def \showDOI       #1{#1}\fi
\ifx \showISBNx    \undefined \def \showISBNx     #1{\unskip}     \fi
\ifx \showISBNxiii \undefined \def \showISBNxiii  #1{\unskip}     \fi
\ifx \showISSN     \undefined \def \showISSN      #1{\unskip}     \fi
\ifx \showLCCN     \undefined \def \showLCCN      #1{\unskip}     \fi
\ifx \shownote     \undefined \def \shownote      #1{#1}          \fi
\ifx \showarticletitle \undefined \def \showarticletitle #1{#1}   \fi
\ifx \showURL      \undefined \def \showURL       {\relax}        \fi
% invisible to TeX
\providecommand\bibfield[2]{#2}
\providecommand\bibinfo[2]{#2}
\providecommand\natexlab[1]{#1}
\providecommand\showeprint[2][]{arXiv:#2}

\bibitem[Antonopoulos(2020)]%
        {antonopoulos2020bankruptcynetwork}
\bibfield{author}{\bibinfo{person}{Angelos Antonopoulos}.} \bibinfo{year}{2020}\natexlab{}.
\newblock \showarticletitle{Bankruptcy problem in network sharing: Fundamentals, applications and challenges}.
\newblock \bibinfo{journal}{\emph{IEEE Wireless Communications}} \bibinfo{volume}{27}, \bibinfo{number}{4} (\bibinfo{year}{2020}), \bibinfo{pages}{81--87}.
\newblock


\bibitem[Balseiro et~al\mbox{.}(2021)]%
        {balseiro2021regularized}
\bibfield{author}{\bibinfo{person}{Santiago Balseiro}, \bibinfo{person}{Haihao Lu}, {and} \bibinfo{person}{Vahab Mirrokni}.} \bibinfo{year}{2021}\natexlab{}.
\newblock \showarticletitle{Regularized online allocation problems: Fairness and beyond}. In \bibinfo{booktitle}{\emph{International Conference on Machine Learning}}. PMLR, \bibinfo{pages}{630--639}.
\newblock


\bibitem[Bardhan and Ashraf(2022)]%
        {bardhan2022more}
\bibfield{author}{\bibinfo{person}{Amit~Kumar Bardhan} {and} \bibinfo{person}{Saad Ashraf}.} \bibinfo{year}{2022}\natexlab{}.
\newblock \showarticletitle{More buyers or more sellers: on marketing resource allocation strategies of competing two-sided platforms}.
\newblock \bibinfo{journal}{\emph{Electronic Commerce Research}} (\bibinfo{year}{2022}), \bibinfo{pages}{1--30}.
\newblock


\bibitem[Beel et~al\mbox{.}(2017)]%
        {beel2017MrDLib}
\bibfield{author}{\bibinfo{person}{Joeran Beel}, \bibinfo{person}{Akiko Aizawa}, \bibinfo{person}{Corinna Breitinger}, {and} \bibinfo{person}{Bela Gipp}.} \bibinfo{year}{2017}\natexlab{}.
\newblock \showarticletitle{Mr. DLib: recommendations-as-a-service (RaaS) for academia}. In \bibinfo{booktitle}{\emph{2017 ACM/IEEE Joint Conference on Digital Libraries (JCDL)}}. IEEE, \bibinfo{pages}{1--2}.
\newblock


\bibitem[Ben-Porat and Torkan(2023)]%
        {ben2023learning}
\bibfield{author}{\bibinfo{person}{Omer Ben-Porat} {and} \bibinfo{person}{Rotem Torkan}.} \bibinfo{year}{2023}\natexlab{}.
\newblock \showarticletitle{Learning with Exposure Constraints in Recommendation Systems}. In \bibinfo{booktitle}{\emph{Proceedings of the ACM Web Conference 2023}}. \bibinfo{pages}{3456--3466}.
\newblock


\bibitem[Biega et~al\mbox{.}(2018)]%
        {biega2018equity}
\bibfield{author}{\bibinfo{person}{Asia~J Biega}, \bibinfo{person}{Krishna~P Gummadi}, {and} \bibinfo{person}{Gerhard Weikum}.} \bibinfo{year}{2018}\natexlab{}.
\newblock \showarticletitle{Equity of attention: Amortizing individual fairness in rankings}. In \bibinfo{booktitle}{\emph{The 41st international acm sigir conference on research \& development in information retrieval}}. \bibinfo{pages}{405--414}.
\newblock


\bibitem[Biswas et~al\mbox{.}(2021)]%
        {biswas2021fairrecplus}
\bibfield{author}{\bibinfo{person}{Arpita Biswas}, \bibinfo{person}{Gourab~K Patro}, \bibinfo{person}{Niloy Ganguly}, \bibinfo{person}{Krishna~P Gummadi}, {and} \bibinfo{person}{Abhijnan Chakraborty}.} \bibinfo{year}{2021}\natexlab{}.
\newblock \showarticletitle{Toward fair recommendation in two-sided platforms}.
\newblock \bibinfo{journal}{\emph{ACM Transactions on the Web (TWEB)}} \bibinfo{volume}{16}, \bibinfo{number}{2} (\bibinfo{year}{2021}), \bibinfo{pages}{1--34}.
\newblock


\bibitem[Boyd and Vandenberghe(2004)]%
        {boyd2004convex}
\bibfield{author}{\bibinfo{person}{Stephen~P Boyd} {and} \bibinfo{person}{Lieven Vandenberghe}.} \bibinfo{year}{2004}\natexlab{}.
\newblock \bibinfo{booktitle}{\emph{Convex optimization}}.
\newblock \bibinfo{publisher}{Cambridge university press}.
\newblock


\bibitem[Cho et~al\mbox{.}(2014)]%
        {cho2014gru}
\bibfield{author}{\bibinfo{person}{Kyunghyun Cho}, \bibinfo{person}{Bart Van~Merri{\"e}nboer}, \bibinfo{person}{Caglar Gulcehre}, \bibinfo{person}{Dzmitry Bahdanau}, \bibinfo{person}{Fethi Bougares}, \bibinfo{person}{Holger Schwenk}, {and} \bibinfo{person}{Yoshua Bengio}.} \bibinfo{year}{2014}\natexlab{}.
\newblock \showarticletitle{Learning phrase representations using RNN encoder-decoder for statistical machine translation}.
\newblock \bibinfo{journal}{\emph{arXiv preprint arXiv:1406.1078}} (\bibinfo{year}{2014}).
\newblock


\bibitem[Claridy(2009)]%
        {claridy2009CUSTOMERSERVICE}
\bibfield{author}{\bibinfo{person}{Sierra Claridy}.} \bibinfo{year}{2009}\natexlab{}.
\newblock \showarticletitle{WHAT IS CUSTOMER SERVICE?}
\newblock \bibinfo{journal}{\emph{Consortium Journal of Hospitality \& Tourism}} \bibinfo{volume}{14}, \bibinfo{number}{1} (\bibinfo{year}{2009}).
\newblock


\bibitem[Cohen and Zhang(2022)]%
        {cohen2022competitiontwo-sidedplatforms}
\bibfield{author}{\bibinfo{person}{Maxime~C Cohen} {and} \bibinfo{person}{Renyu Zhang}.} \bibinfo{year}{2022}\natexlab{}.
\newblock \showarticletitle{Competition and coopetition for two-sided platforms}.
\newblock \bibinfo{journal}{\emph{Production and Operations Management}} \bibinfo{volume}{31}, \bibinfo{number}{5} (\bibinfo{year}{2022}), \bibinfo{pages}{1997--2014}.
\newblock


\bibitem[Curiel et~al\mbox{.}(1987)]%
        {curiel1987bankruptcygames}
\bibfield{author}{\bibinfo{person}{Imma~J Curiel}, \bibinfo{person}{Michael Maschler}, {and} \bibinfo{person}{Stef~H Tijs}.} \bibinfo{year}{1987}\natexlab{}.
\newblock \showarticletitle{Bankruptcy games}.
\newblock \bibinfo{journal}{\emph{Zeitschrift f{\"u}r operations research}}  \bibinfo{volume}{31} (\bibinfo{year}{1987}), \bibinfo{pages}{A143--A159}.
\newblock


\bibitem[Dagan and Volij(1993)]%
        {dagan1993bankruptcycoop}
\bibfield{author}{\bibinfo{person}{Nir Dagan} {and} \bibinfo{person}{Oscar Volij}.} \bibinfo{year}{1993}\natexlab{}.
\newblock \showarticletitle{The bankruptcy problem: a cooperative bargaining approach}.
\newblock \bibinfo{journal}{\emph{Mathematical Social Sciences}} \bibinfo{volume}{26}, \bibinfo{number}{3} (\bibinfo{year}{1993}), \bibinfo{pages}{287--297}.
\newblock


\bibitem[Degefu et~al\mbox{.}(2018)]%
        {degefu2018bankruptcywater}
\bibfield{author}{\bibinfo{person}{Dagmawi~Mulugeta Degefu}, \bibinfo{person}{Weijun He}, \bibinfo{person}{Liang Yuan}, \bibinfo{person}{An Min}, {and} \bibinfo{person}{Qi Zhang}.} \bibinfo{year}{2018}\natexlab{}.
\newblock \showarticletitle{Bankruptcy to surplus: Sharing transboundary river basin’s water under scarcity}.
\newblock \bibinfo{journal}{\emph{Water Resources Management}}  \bibinfo{volume}{32} (\bibinfo{year}{2018}), \bibinfo{pages}{2735--2751}.
\newblock


\bibitem[Donahue et~al\mbox{.}(2015)]%
        {donahue2015lstm}
\bibfield{author}{\bibinfo{person}{Jeffrey Donahue}, \bibinfo{person}{Lisa Anne~Hendricks}, \bibinfo{person}{Sergio Guadarrama}, \bibinfo{person}{Marcus Rohrbach}, \bibinfo{person}{Subhashini Venugopalan}, \bibinfo{person}{Kate Saenko}, {and} \bibinfo{person}{Trevor Darrell}.} \bibinfo{year}{2015}\natexlab{}.
\newblock \showarticletitle{Long-term recurrent convolutional networks for visual recognition and description}. In \bibinfo{booktitle}{\emph{Proceedings of the IEEE conference on computer vision and pattern recognition}}. \bibinfo{pages}{2625--2634}.
\newblock


\bibitem[Duchi et~al\mbox{.}(2011)]%
        {duchi2011subgradient}
\bibfield{author}{\bibinfo{person}{John Duchi}, \bibinfo{person}{Elad Hazan}, {and} \bibinfo{person}{Yoram Singer}.} \bibinfo{year}{2011}\natexlab{}.
\newblock \showarticletitle{Adaptive subgradient methods for online learning and stochastic optimization.}
\newblock \bibinfo{journal}{\emph{Journal of machine learning research}} \bibinfo{volume}{12}, \bibinfo{number}{7} (\bibinfo{year}{2011}).
\newblock


\bibitem[Eisenmann et~al\mbox{.}(2006)]%
        {eisenmann2006strategiesfortwosided}
\bibfield{author}{\bibinfo{person}{Thomas Eisenmann}, \bibinfo{person}{Geoffrey Parker}, \bibinfo{person}{Marshall~W Van~Alstyne}, {et~al\mbox{.}}} \bibinfo{year}{2006}\natexlab{}.
\newblock \showarticletitle{Strategies for two-sided markets}.
\newblock \bibinfo{journal}{\emph{Harvard business review}} \bibinfo{volume}{84}, \bibinfo{number}{10} (\bibinfo{year}{2006}), \bibinfo{pages}{92}.
\newblock


\bibitem[Ertemel and Kumar(2018)]%
        {ertemel2018proportional}
\bibfield{author}{\bibinfo{person}{Sinan Ertemel} {and} \bibinfo{person}{Rajnish Kumar}.} \bibinfo{year}{2018}\natexlab{}.
\newblock \showarticletitle{Proportional rules for state contingent claims}.
\newblock \bibinfo{journal}{\emph{International Journal of Game Theory}}  \bibinfo{volume}{47} (\bibinfo{year}{2018}), \bibinfo{pages}{229--246}.
\newblock


\bibitem[Fleming(2006)]%
        {fleming2006consistency}
\bibfield{author}{\bibinfo{person}{J Fleming}.} \bibinfo{year}{2006}\natexlab{}.
\newblock \showarticletitle{Why consistency is the key to profitable customer service}.
\newblock \bibinfo{journal}{\emph{CUSTOMER MANAGEMENT}} \bibinfo{volume}{14}, \bibinfo{number}{4} (\bibinfo{year}{2006}), \bibinfo{pages}{10}.
\newblock


\bibitem[Gao et~al\mbox{.}(2022)]%
        {gao2022kuairand}
\bibfield{author}{\bibinfo{person}{Chongming Gao}, \bibinfo{person}{Shijun Li}, \bibinfo{person}{Yuan Zhang}, \bibinfo{person}{Jiawei Chen}, \bibinfo{person}{Biao Li}, \bibinfo{person}{Wenqiang Lei}, \bibinfo{person}{Peng Jiang}, {and} \bibinfo{person}{Xiangnan He}.} \bibinfo{year}{2022}\natexlab{}.
\newblock \showarticletitle{KuaiRand: An Unbiased Sequential Recommendation Dataset with Randomly Exposed Videos}. In \bibinfo{booktitle}{\emph{Proceedings of the 31st ACM International Conference on Information \& Knowledge Management}}. \bibinfo{pages}{3953--3957}.
\newblock


\bibitem[Ge et~al\mbox{.}(2020)]%
        {ge2020userriskpref}
\bibfield{author}{\bibinfo{person}{Yingqiang Ge}, \bibinfo{person}{Shuyuan Xu}, \bibinfo{person}{Shuchang Liu}, \bibinfo{person}{Zuohui Fu}, \bibinfo{person}{Fei Sun}, {and} \bibinfo{person}{Yongfeng Zhang}.} \bibinfo{year}{2020}\natexlab{}.
\newblock \showarticletitle{Learning personalized risk preferences for recommendation}. In \bibinfo{booktitle}{\emph{Proceedings of the 43rd International ACM SIGIR Conference on Research and Development in Information Retrieval}}. \bibinfo{pages}{409--418}.
\newblock


\bibitem[Guo et~al\mbox{.}(2019)]%
        {guo2019streaming}
\bibfield{author}{\bibinfo{person}{Lei Guo}, \bibinfo{person}{Hongzhi Yin}, \bibinfo{person}{Qinyong Wang}, \bibinfo{person}{Tong Chen}, \bibinfo{person}{Alexander Zhou}, {and} \bibinfo{person}{Nguyen Quoc Viet~Hung}.} \bibinfo{year}{2019}\natexlab{}.
\newblock \showarticletitle{Streaming session-based recommendation}. In \bibinfo{booktitle}{\emph{Proceedings of the 25th ACM SIGKDD international conference on knowledge discovery \& data mining}}. \bibinfo{pages}{1569--1577}.
\newblock


\bibitem[He et~al\mbox{.}(2020)]%
        {he2020lightgcn}
\bibfield{author}{\bibinfo{person}{Xiangnan He}, \bibinfo{person}{Kuan Deng}, \bibinfo{person}{Xiang Wang}, \bibinfo{person}{Yan Li}, \bibinfo{person}{Yongdong Zhang}, {and} \bibinfo{person}{Meng Wang}.} \bibinfo{year}{2020}\natexlab{}.
\newblock \showarticletitle{Lightgcn: Simplifying and powering graph convolution network for recommendation}. In \bibinfo{booktitle}{\emph{Proceedings of the 43rd International ACM SIGIR conference on research and development in Information Retrieval}}. \bibinfo{pages}{639--648}.
\newblock


\bibitem[He et~al\mbox{.}(2017)]%
        {he2017neuralmf}
\bibfield{author}{\bibinfo{person}{Xiangnan He}, \bibinfo{person}{Lizi Liao}, \bibinfo{person}{Hanwang Zhang}, \bibinfo{person}{Liqiang Nie}, \bibinfo{person}{Xia Hu}, {and} \bibinfo{person}{Tat-Seng Chua}.} \bibinfo{year}{2017}\natexlab{}.
\newblock \showarticletitle{Neural collaborative filtering}. In \bibinfo{booktitle}{\emph{Proceedings of the 26th international conference on world wide web}}. \bibinfo{pages}{173--182}.
\newblock


\bibitem[Hong et~al\mbox{.}(2021)]%
        {hong2021coronavirus}
\bibfield{author}{\bibinfo{person}{Mengdian Hong}, \bibinfo{person}{Ruoying Li}, {and} \bibinfo{person}{Peiran Xie}.} \bibinfo{year}{2021}\natexlab{}.
\newblock \showarticletitle{Construction of Distribution System of Coronavirus Based on Talmud Bankruptcy Distribution}. In \bibinfo{booktitle}{\emph{2021 3rd International Conference on Economic Management and Cultural Industry (ICEMCI 2021)}}. Atlantis Press, \bibinfo{pages}{2089--2097}.
\newblock


\bibitem[Kahneman and Tversky(2013)]%
        {kahneman2013prospecttheory}
\bibfield{author}{\bibinfo{person}{Daniel Kahneman} {and} \bibinfo{person}{Amos Tversky}.} \bibinfo{year}{2013}\natexlab{}.
\newblock \showarticletitle{Prospect theory: An analysis of decision under risk}.
\newblock In \bibinfo{booktitle}{\emph{Handbook of the fundamentals of financial decision making: Part I}}. \bibinfo{publisher}{World Scientific}, \bibinfo{pages}{99--127}.
\newblock


\bibitem[K{\H{o}}szegi and Rabin(2006)]%
        {kHoszegi2006lossaversionuserexp}
\bibfield{author}{\bibinfo{person}{Botond K{\H{o}}szegi} {and} \bibinfo{person}{Matthew Rabin}.} \bibinfo{year}{2006}\natexlab{}.
\newblock \showarticletitle{A model of reference-dependent preferences}.
\newblock \bibinfo{journal}{\emph{The Quarterly Journal of Economics}} \bibinfo{volume}{121}, \bibinfo{number}{4} (\bibinfo{year}{2006}), \bibinfo{pages}{1133--1165}.
\newblock


\bibitem[Leonhardt et~al\mbox{.}(2018)]%
        {leonhardt2018userfairness}
\bibfield{author}{\bibinfo{person}{Jurek Leonhardt}, \bibinfo{person}{Avishek Anand}, {and} \bibinfo{person}{Megha Khosla}.} \bibinfo{year}{2018}\natexlab{}.
\newblock \showarticletitle{User fairness in recommender systems}. In \bibinfo{booktitle}{\emph{Companion Proceedings of the The Web Conference 2018}}. \bibinfo{pages}{101--102}.
\newblock


\bibitem[Li et~al\mbox{.}(2021)]%
        {li2021user}
\bibfield{author}{\bibinfo{person}{Yunqi Li}, \bibinfo{person}{Hanxiong Chen}, \bibinfo{person}{Zuohui Fu}, \bibinfo{person}{Yingqiang Ge}, {and} \bibinfo{person}{Yongfeng Zhang}.} \bibinfo{year}{2021}\natexlab{}.
\newblock \showarticletitle{User-oriented fairness in recommendation}. In \bibinfo{booktitle}{\emph{Proceedings of the Web Conference 2021}}. \bibinfo{pages}{624--632}.
\newblock


\bibitem[Lopes et~al\mbox{.}(2024)]%
        {lopes2024recommendations}
\bibfield{author}{\bibinfo{person}{Ramon Lopes}, \bibinfo{person}{Rodrigo Alves}, \bibinfo{person}{Antoine Ledent}, \bibinfo{person}{Rodrygo~LT Santos}, {and} \bibinfo{person}{Marius Kloft}.} \bibinfo{year}{2024}\natexlab{}.
\newblock \showarticletitle{Recommendations with minimum exposure guarantees: A post-processing framework}.
\newblock \bibinfo{journal}{\emph{Expert Systems with Applications}}  \bibinfo{volume}{236} (\bibinfo{year}{2024}), \bibinfo{pages}{121164}.
\newblock


\bibitem[Lotov and Miettinen(2008)]%
        {lotov2008pareto}
\bibfield{author}{\bibinfo{person}{Alexander~V Lotov} {and} \bibinfo{person}{Kaisa Miettinen}.} \bibinfo{year}{2008}\natexlab{}.
\newblock \showarticletitle{Visualizing the Pareto frontier}.
\newblock In \bibinfo{booktitle}{\emph{Multiobjective optimization: interactive and evolutionary approaches}}. \bibinfo{publisher}{Springer}, \bibinfo{pages}{213--243}.
\newblock


\bibitem[Morik et~al\mbox{.}(2020)]%
        {morik2020controlling}
\bibfield{author}{\bibinfo{person}{Marco Morik}, \bibinfo{person}{Ashudeep Singh}, \bibinfo{person}{Jessica Hong}, {and} \bibinfo{person}{Thorsten Joachims}.} \bibinfo{year}{2020}\natexlab{}.
\newblock \showarticletitle{Controlling fairness and bias in dynamic learning-to-rank}. In \bibinfo{booktitle}{\emph{Proceedings of the 43rd international ACM SIGIR conference on research and development in information retrieval}}. \bibinfo{pages}{429--438}.
\newblock


\bibitem[Naghiaei et~al\mbox{.}(2022)]%
        {naghiaei2022cpfair}
\bibfield{author}{\bibinfo{person}{Mohammadmehdi Naghiaei}, \bibinfo{person}{Hossein~A Rahmani}, {and} \bibinfo{person}{Yashar Deldjoo}.} \bibinfo{year}{2022}\natexlab{}.
\newblock \showarticletitle{Cpfair: Personalized consumer and producer fairness re-ranking for recommender systems}. In \bibinfo{booktitle}{\emph{Proceedings of the 45th International ACM SIGIR Conference on Research and Development in Information Retrieval}}. \bibinfo{pages}{770--779}.
\newblock


\bibitem[O'Neill(1982)]%
        {o1982problem}
\bibfield{author}{\bibinfo{person}{Barry O'Neill}.} \bibinfo{year}{1982}\natexlab{}.
\newblock \showarticletitle{A problem of rights arbitration from the Talmud}.
\newblock \bibinfo{journal}{\emph{Mathematical social sciences}} \bibinfo{volume}{2}, \bibinfo{number}{4} (\bibinfo{year}{1982}), \bibinfo{pages}{345--371}.
\newblock


\bibitem[Patro et~al\mbox{.}(2020)]%
        {patro2020fairrec}
\bibfield{author}{\bibinfo{person}{Gourab~K Patro}, \bibinfo{person}{Arpita Biswas}, \bibinfo{person}{Niloy Ganguly}, \bibinfo{person}{Krishna~P Gummadi}, {and} \bibinfo{person}{Abhijnan Chakraborty}.} \bibinfo{year}{2020}\natexlab{}.
\newblock \showarticletitle{Fairrec: Two-sided fairness for personalized recommendations in two-sided platforms}. In \bibinfo{booktitle}{\emph{Proceedings of the web conference 2020}}. \bibinfo{pages}{1194--1204}.
\newblock


\bibitem[Paudel et~al\mbox{.}(2018)]%
        {paudel2018lossaversion}
\bibfield{author}{\bibinfo{person}{Bibek Paudel}, \bibinfo{person}{Sandro Luck}, {and} \bibinfo{person}{Abraham Bernstein}.} \bibinfo{year}{2018}\natexlab{}.
\newblock \showarticletitle{Loss aversion in recommender systems: Utilizing negative user preference to improve recommendation quality}.
\newblock \bibinfo{journal}{\emph{arXiv preprint arXiv:1812.11422}} (\bibinfo{year}{2018}).
\newblock


\bibitem[Pires and Simon(2015)]%
        {pires2015youtube}
\bibfield{author}{\bibinfo{person}{Karine Pires} {and} \bibinfo{person}{Gwendal Simon}.} \bibinfo{year}{2015}\natexlab{}.
\newblock \showarticletitle{YouTube live and Twitch: a tour of user-generated live streaming systems}. In \bibinfo{booktitle}{\emph{Proceedings of the 6th ACM multimedia systems conference}}. \bibinfo{pages}{225--230}.
\newblock


\bibitem[Qi et~al\mbox{.}(2022)]%
        {qi2022profairrec}
\bibfield{author}{\bibinfo{person}{Tao Qi}, \bibinfo{person}{Fangzhao Wu}, \bibinfo{person}{Chuhan Wu}, \bibinfo{person}{Peijie Sun}, \bibinfo{person}{Le Wu}, \bibinfo{person}{Xiting Wang}, \bibinfo{person}{Yongfeng Huang}, {and} \bibinfo{person}{Xing Xie}.} \bibinfo{year}{2022}\natexlab{}.
\newblock \showarticletitle{Profairrec: Provider fairness-aware news recommendation}. In \bibinfo{booktitle}{\emph{Proceedings of the 45th International ACM SIGIR Conference on Research and Development in Information Retrieval}}. \bibinfo{pages}{1164--1173}.
\newblock


\bibitem[Rendle et~al\mbox{.}(2012)]%
        {rendle2012bpr}
\bibfield{author}{\bibinfo{person}{Steffen Rendle}, \bibinfo{person}{Christoph Freudenthaler}, \bibinfo{person}{Zeno Gantner}, {and} \bibinfo{person}{Lars Schmidt-Thieme}.} \bibinfo{year}{2012}\natexlab{}.
\newblock \showarticletitle{BPR: Bayesian personalized ranking from implicit feedback}.
\newblock \bibinfo{journal}{\emph{arXiv preprint arXiv:1205.2618}} (\bibinfo{year}{2012}).
\newblock


\bibitem[Rochet and Tirole(2004)]%
        {rochet2004two-sided_overview}
\bibfield{author}{\bibinfo{person}{Jean-Charles Rochet} {and} \bibinfo{person}{Jean Tirole}.} \bibinfo{year}{2004}\natexlab{}.
\newblock \showarticletitle{Two-sided markets: An overview}.
\newblock \bibinfo{journal}{\emph{Institut d’Economie Industrielle working paper}} (\bibinfo{year}{2004}), \bibinfo{pages}{1--44}.
\newblock


\bibitem[Rubinstein and Kroese(2016)]%
        {rubinstein2016montecarlo}
\bibfield{author}{\bibinfo{person}{Reuven~Y Rubinstein} {and} \bibinfo{person}{Dirk~P Kroese}.} \bibinfo{year}{2016}\natexlab{}.
\newblock \bibinfo{booktitle}{\emph{Simulation and the Monte Carlo method}}.
\newblock \bibinfo{publisher}{John Wiley \& Sons}.
\newblock


\bibitem[S{\"u}rer et~al\mbox{.}(2018)]%
        {surer2018multistakeholder}
\bibfield{author}{\bibinfo{person}{{\"O}zge S{\"u}rer}, \bibinfo{person}{Robin Burke}, {and} \bibinfo{person}{Edward~C Malthouse}.} \bibinfo{year}{2018}\natexlab{}.
\newblock \showarticletitle{Multistakeholder recommendation with provider constraints}. In \bibinfo{booktitle}{\emph{Proceedings of the 12th ACM Conference on Recommender Systems}}. \bibinfo{pages}{54--62}.
\newblock


\bibitem[Suryahadi et~al\mbox{.}(2003)]%
        {suryahadi2003minimumwage}
\bibfield{author}{\bibinfo{person}{Asep Suryahadi}, \bibinfo{person}{Wenefrida Widyanti}, \bibinfo{person}{Daniel Perwira}, {and} \bibinfo{person}{Sudarno Sumarto}.} \bibinfo{year}{2003}\natexlab{}.
\newblock \showarticletitle{Minimum wage policy and its impact on employment in the urban formal sector}.
\newblock \bibinfo{journal}{\emph{Bulletin of Indonesian economic studies}} \bibinfo{volume}{39}, \bibinfo{number}{1} (\bibinfo{year}{2003}), \bibinfo{pages}{29--50}.
\newblock


\bibitem[Thomson(2003)]%
        {thomson2003axiomatic}
\bibfield{author}{\bibinfo{person}{William Thomson}.} \bibinfo{year}{2003}\natexlab{}.
\newblock \showarticletitle{Axiomatic and game-theoretic analysis of bankruptcy and taxation problems: a survey}.
\newblock \bibinfo{journal}{\emph{Mathematical social sciences}} \bibinfo{volume}{45}, \bibinfo{number}{3} (\bibinfo{year}{2003}), \bibinfo{pages}{249--297}.
\newblock


\bibitem[Thomson(2013)]%
        {thomson2013game}
\bibfield{author}{\bibinfo{person}{William Thomson}.} \bibinfo{year}{2013}\natexlab{}.
\newblock \showarticletitle{Game-theoretic analysis of bankruptcy and taxation problems: Recent advances}.
\newblock \bibinfo{journal}{\emph{International Game Theory Review}} \bibinfo{volume}{15}, \bibinfo{number}{03} (\bibinfo{year}{2013}), \bibinfo{pages}{1340018}.
\newblock


\bibitem[Wang et~al\mbox{.}(2023a)]%
        {wang2023pct}
\bibfield{author}{\bibinfo{person}{Chenyang Wang}, \bibinfo{person}{Yankai Liu}, \bibinfo{person}{Yuanqing Yu}, \bibinfo{person}{Weizhi Ma}, \bibinfo{person}{Min Zhang}, \bibinfo{person}{Yiqun Liu}, \bibinfo{person}{Haitao Zeng}, \bibinfo{person}{Junlan Feng}, {and} \bibinfo{person}{Chao Deng}.} \bibinfo{year}{2023}\natexlab{a}.
\newblock \showarticletitle{Two-sided Calibration for Quality-aware Responsible Recommendation}. In \bibinfo{booktitle}{\emph{Proceedings of the 17th ACM Conference on Recommender Systems}}. \bibinfo{pages}{223--233}.
\newblock


\bibitem[Wang and Joachims(2023)]%
        {wang2023uncertainty}
\bibfield{author}{\bibinfo{person}{Lequn Wang} {and} \bibinfo{person}{Thorsten Joachims}.} \bibinfo{year}{2023}\natexlab{}.
\newblock \showarticletitle{Uncertainty Quantification for Fairness in Two-Stage Recommender Systems}. In \bibinfo{booktitle}{\emph{Proceedings of the Sixteenth ACM International Conference on Web Search and Data Mining}}. \bibinfo{pages}{940--948}.
\newblock


\bibitem[Wang et~al\mbox{.}(2023b)]%
        {wang2023surveyonfairness}
\bibfield{author}{\bibinfo{person}{Yifan Wang}, \bibinfo{person}{Weizhi Ma}, \bibinfo{person}{Min Zhang}, \bibinfo{person}{Yiqun Liu}, {and} \bibinfo{person}{Shaoping Ma}.} \bibinfo{year}{2023}\natexlab{b}.
\newblock \showarticletitle{A survey on the fairness of recommender systems}.
\newblock \bibinfo{journal}{\emph{ACM Transactions on Information Systems}} \bibinfo{volume}{41}, \bibinfo{number}{3} (\bibinfo{year}{2023}), \bibinfo{pages}{1--43}.
\newblock


\bibitem[Wei et~al\mbox{.}(2011)]%
        {wei2011differentupdateinterval}
\bibfield{author}{\bibinfo{person}{Dong Wei}, \bibinfo{person}{Tao Zhou}, \bibinfo{person}{Giulio Cimini}, \bibinfo{person}{Pei Wu}, \bibinfo{person}{Weiping Liu}, {and} \bibinfo{person}{Yi-Cheng Zhang}.} \bibinfo{year}{2011}\natexlab{}.
\newblock \showarticletitle{Effective mechanism for social recommendation of news}.
\newblock \bibinfo{journal}{\emph{Physica A: Statistical Mechanics and its Applications}} \bibinfo{volume}{390}, \bibinfo{number}{11} (\bibinfo{year}{2011}), \bibinfo{pages}{2117--2126}.
\newblock


\bibitem[Wu et~al\mbox{.}(2021)]%
        {wu2021tfrom}
\bibfield{author}{\bibinfo{person}{Yao Wu}, \bibinfo{person}{Jian Cao}, \bibinfo{person}{Guandong Xu}, {and} \bibinfo{person}{Yudong Tan}.} \bibinfo{year}{2021}\natexlab{}.
\newblock \showarticletitle{Tfrom: A two-sided fairness-aware recommendation model for both customers and providers}. In \bibinfo{booktitle}{\emph{Proceedings of the 44th International ACM SIGIR Conference on Research and Development in Information Retrieval}}. \bibinfo{pages}{1013--1022}.
\newblock


\bibitem[Xu et~al\mbox{.}(2023)]%
        {xu2023p}
\bibfield{author}{\bibinfo{person}{Chen Xu}, \bibinfo{person}{Sirui Chen}, \bibinfo{person}{Jun Xu}, \bibinfo{person}{Weiran Shen}, \bibinfo{person}{Xiao Zhang}, \bibinfo{person}{Gang Wang}, {and} \bibinfo{person}{Zhenhua Dong}.} \bibinfo{year}{2023}\natexlab{}.
\newblock \showarticletitle{P-MMF: Provider Max-min Fairness Re-ranking in Recommender System}. In \bibinfo{booktitle}{\emph{Proceedings of the ACM Web Conference 2023}}. \bibinfo{pages}{3701--3711}.
\newblock


\bibitem[Xu et~al\mbox{.}(2024a)]%
        {xu2024fairsync}
\bibfield{author}{\bibinfo{person}{Chen Xu}, \bibinfo{person}{Jun Xu}, \bibinfo{person}{Yiming Ding}, \bibinfo{person}{Xiao Zhang}, {and} \bibinfo{person}{Qi Qi}.} \bibinfo{year}{2024}\natexlab{a}.
\newblock \showarticletitle{FairSync: Ensuring Amortized Group Exposure in Distributed Recommendation Retrieval}.
\newblock \bibinfo{journal}{\emph{arXiv preprint arXiv:2402.10628}} (\bibinfo{year}{2024}).
\newblock


\bibitem[Xu et~al\mbox{.}(2024b)]%
        {xu2024taxation}
\bibfield{author}{\bibinfo{person}{Chen Xu}, \bibinfo{person}{Xiaopeng Ye}, \bibinfo{person}{Wenjie Wang}, \bibinfo{person}{Liang Pang}, \bibinfo{person}{Jun Xu}, {and} \bibinfo{person}{Tat-Seng Chua}.} \bibinfo{year}{2024}\natexlab{b}.
\newblock \showarticletitle{A Taxation Perspective for Fair Re-ranking}. In \bibinfo{booktitle}{\emph{Proceedings of the 47th International ACM SIGIR Conference on Research and Development in Information Retrieval}}. \bibinfo{pages}{1494--1503}.
\newblock


\bibitem[Yang et~al\mbox{.}(2023)]%
        {yang2023vertical}
\bibfield{author}{\bibinfo{person}{Tao Yang}, \bibinfo{person}{Zhichao Xu}, {and} \bibinfo{person}{Qingyao Ai}.} \bibinfo{year}{2023}\natexlab{}.
\newblock \showarticletitle{Vertical Allocation-based Fair Exposure Amortizing in Ranking}. In \bibinfo{booktitle}{\emph{Proceedings of the Annual International ACM SIGIR Conference on Research and Development in Information Retrieval in the Asia Pacific Region}}. \bibinfo{pages}{234--244}.
\newblock


\bibitem[Zanardi and Capra(2011)]%
        {zanardi2011dynamicupdating}
\bibfield{author}{\bibinfo{person}{Valentina Zanardi} {and} \bibinfo{person}{Licia Capra}.} \bibinfo{year}{2011}\natexlab{}.
\newblock \showarticletitle{Dynamic updating of online recommender systems via feed-forward controllers}. In \bibinfo{booktitle}{\emph{Proceedings of the 6th International Symposium on Software Engineering for Adaptive and Self-Managing Systems}}. \bibinfo{pages}{11--19}.
\newblock


\bibitem[Zheng et~al\mbox{.}(2022)]%
        {zheng2022water}
\bibfield{author}{\bibinfo{person}{Yang Zheng}, \bibinfo{person}{Xuefeng Sang}, \bibinfo{person}{Zhiwu Liu}, \bibinfo{person}{Siqi Zhang}, {and} \bibinfo{person}{Pan Liu}.} \bibinfo{year}{2022}\natexlab{}.
\newblock \showarticletitle{Water allocation management under scarcity: a bankruptcy approach}.
\newblock \bibinfo{journal}{\emph{Water Resources Management}} \bibinfo{volume}{36}, \bibinfo{number}{9} (\bibinfo{year}{2022}), \bibinfo{pages}{2891--2912}.
\newblock


\bibitem[Zhou et~al\mbox{.}(2021)]%
        {zhou2021informer}
\bibfield{author}{\bibinfo{person}{Haoyi Zhou}, \bibinfo{person}{Shanghang Zhang}, \bibinfo{person}{Jieqi Peng}, \bibinfo{person}{Shuai Zhang}, \bibinfo{person}{Jianxin Li}, \bibinfo{person}{Hui Xiong}, {and} \bibinfo{person}{Wancai Zhang}.} \bibinfo{year}{2021}\natexlab{}.
\newblock \showarticletitle{Informer: Beyond efficient transformer for long sequence time-series forecasting}. In \bibinfo{booktitle}{\emph{Proceedings of the AAAI conference on artificial intelligence}}, Vol.~\bibinfo{volume}{35}. \bibinfo{pages}{11106--11115}.
\newblock


\end{thebibliography}

\end{document}